\documentclass[aps,prl,reprint,amsmath,amssymb,
superscriptaddress,showpacs,showkeys]{revtex4-1}
\usepackage[T2A]{fontenc}
\usepackage[utf8]{inputenc}
\usepackage{amsmath}    
\usepackage{graphicx}   
\usepackage{verbatim}   
\usepackage{color}      
\usepackage{subfigure}  
\usepackage{dcolumn}    
\usepackage{hyperref}   
\usepackage[warn]{mathtext}    
\usepackage[english,russian]{babel} 
\usepackage{dcolumn}
\usepackage{bm}
\usepackage{amssymb,amsmath}
\usepackage{soulutf8} 
\usepackage{soul} 
\usepackage{color}
\definecolor{vs}{rgb}{0.1,0.4,0.1}                  

\usepackage{verbatim}
\newcommand\wordcount{\verbatiminput{\jobname.sum}}
\usepackage{hyperref}
\begin{document}


\title{Резонансное рассеяние электромагнитных волн малыми металлическими частицами}


\author{М.И. Трибельский}
\email[Corresponding author (replace ``\_at\_" by @):\\]{\mbox{E-mail: mitribelsky\_at\_gmail.com}
}
\homepage[]{https://polly.phys.msu.ru/en/labs/Tribelsky/}
\affiliation{
МГУ, Физический факультет, Москва, Россия}
\affiliation{НИЯУ МИФИ, Инженерно-физический институт биомедицины, Москва, Россия} 
\affiliation{ИОФ РАН, Москва, Россия}
\affiliation{Yamaguchi University, RITS, Yamaguchi, Japan}
\author{A.E. Мирошниченко} 
\affiliation{School of Engineering and Information Technology, UNSW Canberra, Australia}

\date{\today}

\begin{abstract}
Обзор посвящен обсуждению новых (и часто неожиданных) аспектов старой проблемы упругого рассеяния света малыми металлическими частицами, размер которых сопоставим или меньше чем толщина скин-слоя. Основное внимание уделяется выяснению физических основ этих новых аспектов. Показано, что в ряде практически важных случаев рассеяние света такими частицами, несмотря на их малость, может не иметь почти ничего общего с рэлеевским. Детально обсуждается т.н. аномальное рассеяние и поглощение, а также резонансы Фано, в том числе нетрадиционные (связанные с возбуждением продольных электромагнитных колебаний) и пространственные резонансы Фано, наблюдающиеся только в малом телесном угле. Обзор содержит Математическое дополнение, включающее сводку основных результатов теории Ми и обсуждение некоторых общих свойств коэффициентов рассеяния. Помимо чисто академического интереса рассматриваемые в обзоре явления могут найти широкое применение в биологии, медицине, фармацевтике, генной инженерии, визуализации сверхмалых объектов, спектроскопии сверхвысокого разрешения, систем передачи, записи и обработки информации и многих других приложениях и технологиях.
\end{abstract}


\maketitle
%
%

\tableofcontents{}
\section{ Введение}

Рассеяние света малыми частицами имеет давнюю историю. Однако несмотря на то, что точное решение для пространственно-однородной сферической частицы произвольного размера, обладающей произвольным значением диэлектрической проницаемости, было получено Густавом Ми, (а так же независимо Лоренцом и Дебаем) более 100 лет назад~\footnote{Прекрасный исторический обзор этой проблемы, содержится в книге Керкера~\cite{Kerker2013}.} данная задача еще весьма далека от её полного завершения. Дело в том, что в решении Ми поле рассеянной волны представляется в виде бесконечного ряда парциальных волн (дипольной, квадрупольной и т.д.), каждая из которых, в свою очередь, разбивается на сумму, так называемых, электрических и магнитных компонент~\cite{Kerker2013,mie1908,Bohren::1998,Born_Wolf:Optics}. При этом соответствующие эффективности (безразмерные величины, равные отношению парциальных сечений к площади геометрического сечения сферы $\pi R^2$) связаны с комплексными коэффициентами рассеяния $a_n,\; b_n$ довольно простыми, выражениями (подробнее см. ниже).

Однако, как известно, <<дьявол в деталях>> --- явные выражения для самих коэффициентов рассеяния представляют собой громоздкие комбинации специальных функций (т.н., функции Риккати-Бесселя),
см. Математическое дополнение (МД) к настоящему обзору. Их зависимость от различных параметров задачи существенно отличается при различных характерных значениях этих параметров. А это, в свою очередь, приводит к качественному отличию свойств рассеянного излучения в различных областях пространства параметров задачи. Выяснение этих областей и соответствующих свойств рассеянного излучение представляет собой \emph{физическую} (а не формально-математическую, как получение самого точного решения Ми) проблему. И именно \emph{эта} проблема все еще преподносит пытливому исследователю неожиданные и весьма важные результаты.

В последнее время интерес к ней особенно возрос в связи с громадным прогрессом в нанотехнологиях и, как следствие этого, широким применениям результатов исследования задачи Ми и ее обобщений в самых разных областях --- от телекоммуникации до диагностики и лечения онкологических и других заболеваний~\cite{Eaton:VA:1984,Krasnok:OE:2012,Chattaraj:JOSA:2016,Chaabani:N:2019,Staude:P:2019}. Соответственно выглядит и число публикаций по данной теме. Поиск в {\it Google Scholar} по ключевым словам дает без малого полтора миллиона (!) монографий, статей и программных продуктов.

Разумеется, нет никакой возможности представить в одной статье обзор всей этой массы материала. Авторы настоящей публикации и не ставят перед собой такой невыполнимой задачи. Тем более, что по данной теме уже существует большое число как прекрасных монографий (см., например,~\cite{Wriedt:PPSC:1998,Kerker2013,Bohren::1998,Born_Wolf:Optics}), так и замечательных обзоров ~\cite{Kerker:AST:1982,Kelly:JPC:2003,Wriedt:Springer:2012,Kuznetsov:S:2016,Tzarouchis:AS:2018}. Однако поскольку развитие этого направления происходит все ускоряющимися темпами, даже наиболее свежие обзорные статьи устаревают уже к моменту их выхода в свет. {По этой причине мы решили сконцентрироваться на некоторых последних достижениях в этой области, которые ближе всего к собственным исследованиям авторов. Но даже и здесь мы не претендуем на полноту, ограничившись только несколькими, с нашей точки зрения, наиболее интересными вопросами рассеяния и поглощения электромагнитного излучения малыми металлическими частицами, и оставляя обсуждение диэлектрических частиц для будущего обзора, если таковой когда-либо будет написан.} При этом, мы старались основное внимание уделить физике обсуждаемых явлений, ограничившись минимальным количеством простых формул. Все более сложные математические выкладки вынесены в МД. Обсуждение эксперимента также ограничено самым необходимым минимумом.

{Однако когда, следуя означенным принципам, мы приступили к написанию настоящего обзора, выяснилось, что и эти ограничения все еще недостаточны для удержания объема обзора в разумных пределах. Поэтому их пришлось ужесточить. В частности, из обзора пришлось исключить весьма интересные и важные для приложений проблемы оптических свойств как метаматериалов (искусственных, специальным образом структурированных сред), так и <<метамолекул>> --- нескольких наночастиц разделенных промежутками меньшими или сопоставимыми с длиной волны падающего излучения, и образующими по отношению к падающей волне определнную пространственную конфигурацию. Читателя, интересующегося этими вопросами, нам приходится адресовать к другой оригинальной и обзорной литературе, см., например,~\cite{AGRANOVICH20091,Merlin1693,Klimov:Nanoplasmonika,Lukyanchuk:2010bt,Rahmani2013}. То, что осталось после всех ограничений --- это эффекты, возникающие при рассеянии света уединенной сферической металлической частицей или сферически-симметричной частицей типа ядро-оболочка, каковые и будут обсуждены достаточно подробно.}

Поскольку различные читатели могут иметь различные области интересов, мы, сохраняя единство изложения и его логическую последовательность, постарались сделать каждый раздел обзора самодостаточным. Это позволит при желании ограничится чтением только отдельных разделов обзора.

Всюду, где это не будет оговорено особо, мы будем рассматривать рассеяние плоской линейно поляризованной волны с временн\'{о}й зависимостью полей, пропорциональной $\exp[-i\omega t]$. Поскольку плоские волны являются полной системой функций, по ним можно разложить произвольную падающую волну, поэтому такое рассмотрение не ограничивает общности обсуждаемых результатов.

Мы будем пользоваться Гауссовой системой единиц. Кроме того, все вещества предполагаются немагнитными, так что магнитная проницаемость $\mu$ всегда считается равной единице. Рассматривается только пассивное рассеяние. Рассеяние активными частицами, имеющими инверсную населенность, и связанные с ним лазерные эффекты лежат вне рамок данной работы. {Подчеркнем, что хотя всюду в настоящем обзоре мы говорим о рассеянии электромагнитных волн, приведенные результаты легко переносятся на случай рассеяния волн другой природы, например, акустических.}

\section{Плазмонные резонансы}
\subsection{Рэлеевское рассеяние и область его применимости}

То, что субволновая частица рассеивает электромагнитное излучение как электрический диполь хорошо известно со времени фундаментальных работ лорда Рэлея~\cite{lord1871light,Rayleigh_Book}, опубликованных 150 лет тому назад~(!), в справедливости которых мы убеждаемся, любуясь голубым небом или багровым закатом. Впоследствии их результаты блестяще подтверждались бесчисленными экспериментальными и теоретическими работами. Казалось бы, что вопрос решен раз и навсегда, и ничего нового здесь никогда уже не может быть обнаружено.

Но,... <<never say never>>. Давайте кратко воспроизведем и проанализируем рассуждения, лежащие в основе знаменитой формулы для сечения рэлеевского рассеяния для сферы радиуса $R$~\cite{LL_8}:

\begin{equation}\label{eq:sigma_Ray}
{\sigma _{\rm sca}} = \frac{8\pi\varepsilon^2_{\rm out}}{{3}}\left(\frac{\omega}{c}\right)^4\!R^6{\left| {\frac{{\varepsilon  - 1}}{{\varepsilon  + 2}}} \right|^2}\equiv \frac{{8\pi {x^6}}}{{3{k^2}}}{\left| {\frac{{\varepsilon  - 1}}{{\varepsilon  + 2}}} \right|^2}
\end{equation}
Здесь $\varepsilon \equiv \varepsilon_1/\varepsilon_{\rm out}$ --- диэлектрическая проницаемость материала сферы, нормированная на соответствующее значение для окружающей сферу среды, а величина $x$ (параметр размера) равна  $kR$, где $k=\omega m_{\rm out}/c$ --- волновое число падающей волны, $m_{\rm out} \equiv \sqrt{\varepsilon_{\rm out}}$ --- коэффициент преломления окружающей среды, а $c$ --- скорость света в вакууме. Параметр размера является чисто действительной величиной, т.к. рассматриваемая постановка задачи (рассеяния электромагнитных волн в неограниченной внешней среде, когда падающая волна приходит из бесконечности, а рассеянное излучение уходит на бесконечность) возможна только при чисто действительном \mbox{значении $m_{\rm out}$.} {Отметим, что часто наряду с размерной величиной $\sigma _{\rm sca}$ бывает удобно характеризовать интенсивность рассеяния безразмерной эффективностью рассеяния: \mbox{$Q_{\rm sca} = \sigma _{\rm sca}/(\pi R^2)$.}}

Если форма рассеивающей свет частицы близка к сферической, а ее радиус значительно меньше длины волны излучения, в том числе и в материале сферы, то в каждый момент времени электрическое поле падающей на частицу электромагнитной волны можно считать пространственно-однородным. В этом случае из симметрии задачи следует, что поле внутри частицы также пространственно-однородно, хотя значение вектора $\mathbf{E}$ внутри частицы определяется $\varepsilon$ и отлично от соответствующего значения в падающей волне~\cite{LL_8}. Однородное поле создает в частице наведенный дипольный момент, величина которого пропорциональна объему частицы, т.е. $R^3$, и вместе с полем падающей волны осциллирует во времени с частотой $\omega$. Что касается магнитной поляризуемости, то при $\mu$ близком к единице ей можно пренебречь. Интенсивность излучения электрического диполя пропорциональна квадрату модуля второй производной по времени от величины дипольного момента~\cite{LL_2}, т.е. $\omega^4 R^6$, что совпадает с \mbox{зависимостью \eqref{eq:sigma_Ray}.}

Что не учтено в этих рассуждениях? Помимо $\omega^4 R^6$ в формуле \eqref{eq:sigma_Ray} имеется множитель \mbox{$|(\varepsilon - 1)/(\varepsilon + 2)|^2$}, описывающий поляризуемость частицы пространственно-однородным электрическим полем~\cite{LL_8}. {Заметим теперь, что для хороших металлов в оптической области спектра $\varepsilon <0$, а знаменатель \eqref{eq:sigma_Ray} обращается в ноль при $\varepsilon = -2$. Мы приходим к нефизическому результату --- расходимости сечения рассеяния. Подчеркнем, во избежание недоразумений, что хотя при  $\varepsilon <0$ поле в металле экспоненциально затухает, это не противоречит нашему предположению о наличии внутри рассеивающей частицы однородного поля --- как уже отмечалось выше, предполагается, что размер частицы много меньше характерной длины затухания поля (толщины скин-слоя).}

Учет того, что чисто действительное $\varepsilon$ есть идеализированный случай полного отсутствия диссипации, и что в реальности $\varepsilon$ всегда имеет отличную от нуля мнимую часть, которая описывает диссипативные процессы не спасает положения. Действительно, при наличии у $\varepsilon$ ненулевой мнимой части приводит к  тому, что знаменатель \eqref{eq:sigma_Ray} не может обратиться в ноль. Однако, не существует никаких фундаментальных ограничений снизу на величину Im$\,\varepsilon$, кроме того, что для обеспечения диссипации она должна быть положительна, а в силу принципа причинности удовлетворять соотношению Крамерса-Кронига. Возможны случаи когда при некоторых частотах величина Im$\,\varepsilon$ оказывается чрезвычайно мала. {Если эта область малости Im$\,\varepsilon$ совпадает с областью, в которой Re$\,\varepsilon \approx -2$, то} может статься что согласно формуле \eqref{eq:sigma_Ray} шарик радиусом в несколько нанометров имеет сечение рассеяния в квадратный километр, что противоречит элементарному здравому смыслу. Говоря про квадратный километр, мы, конечно, утрируем, но ясно, что в окрестности $\varepsilon = -2$ рэлеевское приближение может приводить к абсурдным результатам, т.е. перестает работать, а формула \eqref{eq:sigma_Ray} требует корректировки.

Подойдем к этому важному вопросу с другой стороны. Микроскопически, колебания дипольного момента связаны с коллективными колебаниями свободных и/или связанных зарядов в веществе рассеиваемой частицы. Как и всякие колебания они имеют собственные частоты. Когда частота внешнего воздействия становиться равной собственной частоте колебаний, при отсутствии диссипации их амплитуда расходится. Очевидно, что обсуждаемая расходимость выражения \eqref{eq:sigma_Ray} при  $\varepsilon = -2$ как раз и означает такое резонансное возбуждение. Это хорошо согласуется с известным в теории рассеяния фактом расходимости сечения упругого рассеяния в точках резонансов~\cite{LL_3}.

Поскольку, {как уже отмечалось,} отрицательные значения $\varepsilon$ в видимой области спектра обычно соответствуют металлам, будем, для определенности, говорить о свободных электронах. Отмеченные выше колебания электронной плазмы называются локализованными плазмонами. Таким образом, при рассеянии света малой металлической сферой происходит перекачка энергии из падающей электромагнитной волны в локализованные плазмоны. Однако в отсутствии диссипации все процессы обратимы. Если существует прямой процесс, то должен существовать и обратный --- трансформация локализованных плазмонов в распространяющуюся от частицы (т.е. рассеянную) электромагнитную волну.

Процесс этот хорошо известен и носит название радиационного затухания плазмона. При этом, если поток энергии в прямом процессе определяется амплитудой падающей волны, то в обратном процессе он определяется амплитудой плазмона. Обычно радиационное затухание весьма мало, см. ниже. Тем не менее ясно, что как бы ни была мала эффективность обратной трансформации, при неограниченно росте амплитуды плазмона в конце концов она достигает такой величины, что оба процесса приходят в динамическое равновесие, и дальнейший рост амплитуды плазмона (а следовательно и мощности рассеянного излучения) прекращается.

Важно, что в силу малости радиационного затухания локализованный плазмон обладает большим временем жизни. Если в какой-то момент времени скачком отключить падающее излучение, то накопленная в локализованном плазмоне энергия не излучится мгновенно, а будет убывать по экспоненциальному закону с показателем равным времени жизни плазмона. Поэтому при динамическом равновесии, поток энергии при прямой трансформации падающей электромагнитной волны в локализованные плазмоны компенсируется обратным потоком радиационного распада плазмонов, возбужденных \emph{в предыдущие моменты времени}. Иными словами, существует определенный \emph{эффект запаздывания\/} между прямой и обратной трансформациями.

Теперь становится ясно, чт\'{о} именно не учитывает формула Рэлея. Она основана на описании поляризации частицы \emph{стационарным\/} электрическим полем. Такое описание в принципе не может учесть эффекты запаздывания, т.е. радиационное затухание. Вдали от плазмонного резонанса этим затуханием действительно можно пренебречь в силу его малости, что и приводит к формуле \eqref{eq:sigma_Ray}. Однако в окрестности резонанса это пренебрежение становится недопустимым, т.к. при малой диссипации именно радиационное затухание ограничивает величину сечения рассеяния.

\subsection{Аномальное рассеяние}
\subsubsection{Недиссипативный предел}

Что же нужно сделать, чтобы правильно учесть радиационное затухание? В том то и дело, что очень мало. У нас же есть точное решение задачи~\cite{Kerker2013,mie1908,Bohren::1998}, см.  также МД. Поскольку оно \emph{точное\/}, радиационное затухание включено в это решение автоматически. Единственное, что остается сделать --- выделить эффект радиационного затухания в явном виде. Для этого нужно преобразовать громоздкие формулы Ми, воспользовавшись малостью $x$.

Сводка результатов точного решения Ми представлена в МД. В основном тексте обзора мы приведем только самые необходимые выражения. В этом решении поле рассеянной волны представляется в виде мультипольного разложения --- бесконечного ряда по сферическим гармоникам (дипольной, квадрупольной, октупольной и т.д.). Каждая гармоника (парциальная мода) представляет собой сумму двух независимых компонент --- электрической и магнитной. Комплексные амплитуды первой и второй (\emph{коэффициенты рассеяния}) принято обозначать $a_n$ и $b_n$, соответственно, где $n=1,\,2,\,3,\ldots$ --- номер гармоники. В общем случае коэффициенты рассеяния представляют собой довольно громоздкие выражения, записанные чрез функции Риккати-Бесселя от $x$ и $mx$. Здесь $m \equiv \sqrt{\varepsilon}$ --- относительный коэффициент преломления сферы (вообще говоря, комплексный). Однако можно показать (см. МД), что из общих свойств точного решения следует, что
\begin{equation}\label{eq:an_FG}
  a_n=\frac{F^{(a)}_n}{F^{(a)}_n+iG^{(a)}_n}.
\end{equation}
Аналогичное выражение справедливо для $b_n$, разумеется, с другими чем для $a_n$ значениями $F$ и $G$. При этом, при отсутствии диссипации (Im$\,\varepsilon =0$) функции $F$ и $G$ являются чисто действительными.

Полное сечение рассеяния представляется в виде бесконечного ряда парциальных сечений~\cite{bohren1983can}
\begin{equation}\label{eq:sigma_sph_n}
  \sigma_{\rm sca}^{(n)}  = \frac{{2\left( {2n + 1}\right)\pi}}{{k^2}}\left( {|a_n  |^2  + |b_n  |^2 } \right),
\end{equation}
каждое из которых связано с соответствующей парциальной модой.

Раскладывая функции $F$ и $G$ в ряд по малым $x$, легко убедиться, что при $|mx| \ll 1$, выражение для $b_n$ не имеет резонансного знаменателя, а само $b_n$ оказывается порядка $x^{2n+3}$. Что же касается $a_n$, то для него после сокращения общего множителя в числителе и знаменателе выражения \eqref{eq:an_FG}  получаем:
\begin{eqnarray}
  F^{(a)}_n &=&  x^{2n+1}\left\{\frac{(n+1)(\varepsilon-1)}{[(2n+1)!!]^2}+O(x^2)\right\}, \label{eq:F_an_x<<1} \\
  G^{(a)}_n &=& \frac{n}{2n+1}\left(\varepsilon+\frac{n+1}{n}+O(x^2)\right), \label{eq:G_an_x<<1}
\end{eqnarray}

{Казалось бы, что поскольку $F_n$ имеет высокий порядок малости по $x$, а главный член в разложении $G_n$ по степеням $x$ вообще от $x$ не зависит, то либо в знаменателе выражения \eqref{eq:an_FG} мы всегда можем пренебречь $F_n$ по сравнению с $G_n$, либо необходимо учесть в правой части \eqref{eq:G_an_x<<1} члены до $x^{2n+1}$ включительно. Однако в силу того, что в выражении \eqref{eq:an_FG} функция $G_n$ входит с множителем $i$, этот вопрос оказывается несколько более сложным. Рассмотрим его подробнее.}

Если $(\varepsilon +1+1/n) \gg x^{2n+1}$, то $F$ в знаменателе \eqref{eq:an_FG} действительно можно пренебречь по сравнению с $G$. В этом случае все $a_n$ становятся чисто мнимыми (т.е. рассеянное поле сдвинуто по фазе относительно поля падающей волны на $\pm \pi/2$), а их модуль оказывается порядка $x^{2n+1}$. Подстановка соответствующих выражений в \eqref{eq:sigma_sph_n} дает значение парциального сечения рассеяния, являющегося обобщением формулы \eqref{eq:sigma_Ray} на случай произвольного $n$. В частности, при $n=1$ получаем в точности \eqref{eq:sigma_Ray}. При произвольном $n$ соответствующее парциальное сечение оказывается порядка $x^{2(2n+1)}$. Это обстоятельство вместе с отмеченной малостью $b_n$ по сравнению с $a_n$ приводит к тому, что из всего бесконечного ряда основной вклад в полное сечение рассеяние вносит именно электрический дипольный момент, в полном соответствии с проведенным выше обсуждением формулы \eqref{eq:sigma_Ray}.

{Здесь уместно сделать некоторое общее замечание относительно минимально необходимой в рассматриваемом круге задач точности приближенных вычислений. Помимо $\sigma_{\rm sca}$ важным параметром, описывающим процесс рассеяния является сечение экстинкции $\sigma_{\rm ext} \equiv \sigma_{\rm sca}+ \sigma_{\rm abs}$, где $\sigma_{\rm abs}$ --- это сечение поглощения, характеризующее мощность, поглащенную рассеивающим излучение объектом за счет необратимых, диссипативных процессов. Для сферической частицы парциальное сечение экстинкции определяется следующей формулой~\cite{bohren1983can}:
\begin{equation}\label{eq:sigma_ext_sph_n}
  \sigma_{\rm ext}^{(n)}  = \frac{{2\left( {2n + 1}\right)\pi}}{{k^2}}\left( {\rm Re}\,a_n  + {\rm Re}\,b_n  \right),
\end{equation}
В недиссипативном пределе  $\sigma_{\rm abs}=0$, и, следовательно,  $\sigma_{\rm ext} = \sigma_{\rm sca}$. Однако если, в соответствии со сказанным ранее, ограничиться в выражении \eqref{eq:an_FG} главным приближением по малому $F$, то реальная часть от получившегося чисто мнимого выражения равна нулю, между тем как его модуль, очевидно, нулю не равняется, и равенство $\sigma_{\rm ext} = \sigma_{\rm sca}$ не выполняется. Парадокс легко разрешается, если учесть, что при приближенных вычислениях обращение в ноль главного члена означает только то, что необходимо учесть следующий член разложения. В выражении \eqref{eq:sigma_sph_n} для парциального сечения рассеяния главный член ненулевой и члены более высокого порядка можно отбросить. В выражении же \eqref{eq:sigma_ext_sph_n} главный член обнуляется. Тогда, учитывая следующий член разложения правой части \eqref{eq:an_FG} по малому $F$, получаем \mbox{$a \approx -i(F/G) + (F/G)^2$}, и равенство $\sigma_{\rm ext} = \sigma_{\rm sca}$ восстанавливается.
}

{Отметим также, что, рэлеевское приближение в общеупотребительном смысле означает не только пренебрежение в знаменателе выражения для коэффициентов рассеяния $F$ по сравнению с $G$, но и пренебрежение вкладами в сечение рассеяния, поглощения или экстинкции всех парциальных мод за исключением дипольной. Нам, однако, будет удобно пользоваться расширенным толкованием этого термина, означающим пренебрежение $F$ по сравнению с $G$, но не включающего в себя пренебрежение высшими мультиполями. Это позволяет применять его не только для дипольной моды, но также и для любого парциального сечения при $n>1$.}

Обсудим теперь, что происходит  при приближении $\varepsilon$ к $-(n+1)/n$~\cite{Tribelsky:JETP:1984,Tribelsky:2006jf}. Обсуждение начнем с недиссипативного предела чисто действительного $\varepsilon$. Поскольку нас интересует малая окрестность точки  \mbox{$\varepsilon = -(n+1)/n$}, естественно ввести \mbox{$\delta\varepsilon \equiv \varepsilon + (n+1)/n$}. Тогда в главном по малому $\delta\varepsilon$ приближении выражение для $a_n$ приобретает вид
\begin{equation}\label{eq:an&gamma_n_x<<1}
  a_n \approx \frac{i\gamma_n/2}{\delta\varepsilon + i(\gamma_n/2)};\; \gamma_n = \frac{2(n+1)x^{2n+1}}{[n(2n-1)!!]^2}.
\end{equation}
При фиксированном $x$ такая зависимость приводит для $|a_n|^2$ как функции $\delta\varepsilon$ к лоренцевскому профилю с шириной линии равной $\gamma_n$, а при $\delta\varepsilon \gg \gamma_n$ --- к формуле
\begin{equation}\label{eq:an_x<<1_de<<gamma}
  a_n \approx \frac{i\gamma_n}{2\delta\varepsilon},
\end{equation}
что при $n=1$ (дипольное приближение) сводится к формуле Рэлея \eqref{eq:sigma_Ray}.
{Важно, что конечность $\gamma_n$ существенно связана с конечностью $x$. При переходе к внешнему пространственно-однородному полю длина волны падающего излучения стремится к бесконечности, а $x$ (и вместе с ним $\gamma_n$) стремится к нулю. Иными словами, конечная ширина линии возникает только при учете отклонения поля падающей волны от пространственно-однородного --- т.е., при учете эффектов запаздывания, что полностью соответствует качественным рассуждениям о радиационном затухании, приведенным выше при обсуждении формулы Рэлея.}

{Следует также подчеркнуть, что при $\delta\varepsilon=0$ и $\gamma_n=0$ в выражении \eqref{eq:an&gamma_n_x<<1} для $a_n$ возникает неопределенность типа 0/0. Эту неопределенность нельзя раскрыть --- в точке $\delta\varepsilon=0$ и $\gamma_n=0$ правая часть \eqref{eq:an&gamma_n_x<<1} не имеет предела, а сама точка является аналогом существенно особых точек аналитических функций. Особенность устраняется при учете любого, сколь угодно малого, но конечного диссипативного затухания, см. подробнее ниже, а также МД и работу~\cite{Brynkin2019}.}

{Здесь важно отметить, что поскольку в общем случае $F_n$ и $G_n$ являются функциями двух переменных $x$ и $mx$, т.е., фактически, $x$ и $\varepsilon$, то при чисто действительном $\varepsilon$ условие резонанса $G^{(a)}_n=0$ задает на плоскости  $x,\,\varepsilon$ определенную резонансную кривую $\varepsilon = \varepsilon^{\rm res}(x)$. В каждой точке этой кривой $a_n=1$. Обсуждавшаяся же выше точка \mbox{$x=0,\;\varepsilon=-(n+1)/n$}, есть всего лишь начальная точка данной кривой, из которой она выходит при $x=0$. }

{При малом, но конечном $x$ под $\delta\varepsilon$ в выражении \eqref{eq:an&gamma_n_x<<1} следует понимать разность $\varepsilon - \varepsilon^{\rm res}(x)$. Аналогично, и в выражении \eqref{eq:F_an_x<<1} следует положить  $\varepsilon = \varepsilon^{\rm res}(x)$. Однако расчеты показывают, что разность между $\varepsilon^{\rm res}(x)$ и ее значением при $x=0$ оказывается порядка $x^2$~\cite{Tribelsky:2006jf}, см. \eqref{eq:G_an_x<<1}, что дает лишь малые поправки к выражению \eqref{eq:an_x<<1_de<<gamma} для $\gamma_n$. Таким образом, в главном приближении при малом, но конечном $x$ ни положение точки резонанса на оси $\varepsilon$, ни ширина резонансной линии (по $x$) не меняются. Это и является оправданием для пренебрежения членами более высокого порядка в выражениях \eqref{eq:F_an_x<<1}, \eqref{eq:G_an_x<<1}.}

Для применения этого результата к реальной физической ситуации следует от независимых переменных \mbox{$x,\;\varepsilon$} перейти к переменным \mbox{$R,\;\omega$} и рассмотреть форму линии, как функцию $\omega$ при фиксированном $R$. Это легко сделать, раскладывая $\varepsilon(\omega)$ в ряд в точке $\omega = \omega_n^{\rm res}$, где \mbox{$\varepsilon(\omega_n^{\rm res})=\varepsilon^{\rm res}(x)$}. При этом \mbox{$\delta\varepsilon \rightarrow (d\varepsilon/d\omega)_{\omega_n^{\rm res}}(\omega-\omega_n^{\rm res})$.} Что же касается самого $x$, то его в главном приближении можно представить как \mbox{$k(\omega_n^{\rm res})R \equiv m_{\rm out}\omega_n^{\rm res}R/c$.}

Согласно формуле \eqref{eq:sigma_sph_n}, тот же лоренцевский профиль сохраняется и для парциального сечения. В точке резонанса, соответствующей максимуму данного профиля, $a_n=1$, а парциальное сечение оказывается равным
\begin{equation}\label{eq:sigma_sca_n_max}
   \sigma_{\rm sca\,max}^{(n)}  = \frac{{2(2n + 1)\pi}}{{k^2}}.
\end{equation}

{Другой важной особенностью такого резонансного рассеяния является зависимость компонент напряженностей электрического и магнитного полей парциальной рассеянной волны в непосредственной окрестности поверхности частицы  от ее размера. В рэлеевском случае они оцениваются выражениями \mbox{$E^{{\rm \,sca\,}(n)}_{r,\theta,\varphi} \sim x^{n-1}$}, \mbox{$H^{{\rm \,sca\,}(n)}_{\theta,\varphi} \sim x^{n}$}, \mbox{$H^{{\rm \,sca\,}(n)}_{r} \sim x^{n+1}$}, т.е. остаются \emph{регулярными\/} при \mbox{$x \rightarrow 0$}. При резонансном рассеянии при перемещении в плоскости $\varepsilon, x$ вдоль резонансной кривой в направлении $x=0$ эти же величины оцениваются как \mbox{$E^{{\rm\,sca\,}(n)}_{r,\theta,\varphi} \sim x^{-(n+2)}$}, \mbox{$H^{{\rm \,sca\,}(n)}_{\theta,\varphi} \sim x^{-(n+1)}$}, \mbox{$H^{{\rm \,sca\,}(n)}_{r} \sim x^{n+1}$}, т.е. за исключением радиальной компоненты магнитного поля оказываются \emph{сингулярны\/} по $x$ и \emph{возрастают\/} при уменьшении размера рассеивающей частицы~\cite{Tribelsky:2006jf,Tribelsky:2011ir}.} Это свойство есть прямое следствие независимости $ \sigma_{\rm sca\,max}^{(n)}$ от $x$ --- чтобы обеспечить неизменную мощность рассеянного излучения при уменьшении размера излучающей области требуется увеличение амплитуды поля в этой области.

Казалось бы, что отмеченная независимости $ \sigma_{\rm sca\,max}^{(n)}$ от $x$ приводит к парадоксу --- частица нулевого размера, т.е. в реальности \emph{несуществующая\/} по-прежнему имеет конечное сечение рассеяния, что является очевидной нелепостью. Мы разрешим этот парадокс чуть позже, оставляя читателю время самому найти ответ на вопрос, что именно привело к нелепости при анализе решения, которое является \emph{точным\/} и следовательно применимо в той же мере, в какой применимы сами уравнения Максвелла, в справедливости которых никто, конечно, не сомневается. Пока же, продолжим обсуждения необычных свойств резонансного рассеяния.

Как было качественно показано при обсуждении формулы \eqref{eq:sigma_Ray} и количественно --- при обсуждении точного решения Ми, основной вклад в рэлеевское рассеяние вносит дипольная мода. Парциальные сечения высших мультиполей убывают как \mbox{$x^{2(2n+1)}$,} см. \mbox{\eqref{eq:an_FG}--\eqref{eq:G_an_x<<1}}. Напротив, при резонансном рассеянии парциальные сечения \emph{возрастают\/} с ростом $n$ как \mbox{$2(2n+1)$,} см. \eqref{eq:sigma_sca_n_max}, т.е. резонансное значение парциального сечения дипольной моды меньше, чем квадрупольной, которое, в свою очередь, меньше октупольной и т.д. --- таким образом, при резонансном рассеянии возникает \emph{обратная иерархия резонансов}~\cite{Tribelsky:JETP:1984,Libenson2005}. Нужно подчеркнуть, что поскольку резонансные значения $\varepsilon$ для каждого значения $n$ различны ($\varepsilon^{\rm res} \approx -(n+1)/n$), а резонансные линии узкие, перекрытие резонансов разных порядков не происходит, и этот эффект не влияет на сходимость мультипольного разложения для сферы.

{Следует также иметь в виду, что каждый мультиполь имеет свое резонансное значение $\varepsilon$, а следовательно и свою резонансную частоту. Поэтому, когда мы говорим об обратной иерархии резонансов, мы имеем в виду, сравнение сечений при \emph{различных\/} значениях $\omega$ --- своих для каждого из резонансов. При этом, пропорциональность парциальных сечений \mbox{$2(2n+1)$} сохраняется только в том случае, если остается неизменным $k$, см. \eqref{eq:sigma_sca_n_max}. Поскольку сравнение происходит на разных частотах, то неизменность $k$ можно сохранить только сравнивая резонансные сечения для сфер с разными размерами. Такое сравнение, мягко говоря, выглядит несколько странным. Поэтому желательно понять, как зависит сечение рассеяния для одной и той же частицы, когда ее размер фиксирован, а единственным изменяющимся параметром является частота падающего излучения.}

{Как хорошо известно для рэлеевского рассеяния, если пренебречь дисперсией диэлектрической проницаемости, то при фиксированном $R$ интенсивность сечения рассеяния есть непрерывная функция частоты и возрастает с ее ростом как $\omega^4$, а для более высоких мультиполей, как $\omega^{4n}$, см. \mbox{\eqref{eq:sigma_sph_n}, \eqref{eq:F_an_x<<1}.}}
{Что же касается резонансного рассеяния, то в силу узости резонансных линий это рассеяние сосредоточено в малых окрестностях резонансных частот, поэтому зависимость его сечения от частоты дискретная и определяется спектром резонансных частот.}

{В целом частотная зависимость выглядит следующим образом. Имеется относительно широкая линия дипольного резонанса, крылья которой выходят на рэлеевскую зависимость \eqref{eq:sigma_Ray}. На этих крыльях, как на пьедестале, располагаются узкие интенсивные пики, соответствующие резонансному рассеянию, ширина которых уменьшается по мере роста $n$. Детали такой картины определяются конкретным видом дисперсионной зависимости $\varepsilon(\omega)$.}

\begin{figure}
  \centering
  \includegraphics[width=\columnwidth]{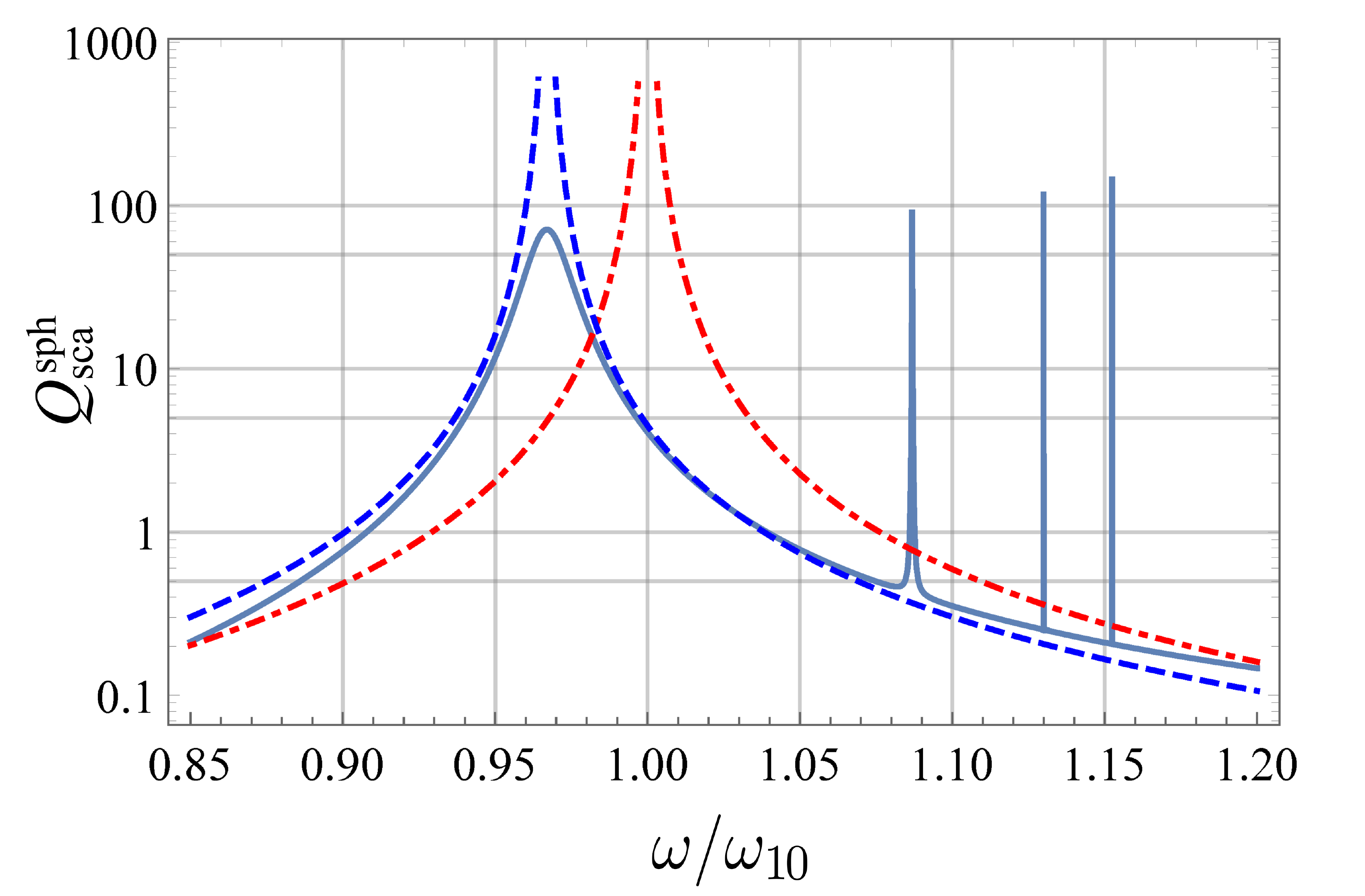}
  \caption{{Сплошная линия --- зависимость эффективности рассеяния сферой с $\varepsilon(\omega)$, задаваемой формулой Друде \eqref{eq:epsilon_Drude}. Расчет на основании точного решения Ми при \mbox{$x_{10}=0,3$.} Штрих-пунктирная линия --- та же величина, рассчитанная по формуле Рэлея \eqref{eq:sigma_Ray}. Пунктирная линия --- дипольная парциальная эффективность, рассчитанная в соответствии с формулой \eqref{eq:an_x<<1_de<<gamma}, где $\delta\varepsilon$ отсчитывается от центра линии дипольного резонанса при $x_{10}=0,3$, т.е. то же рэлеевское рассеяние, но с учетом смещения резонансной частоты за счет конечности размера частицы. Это смещение по отношению к точке к $\omega=\omega_{10}$ хорошо видно на рисунке, так же как обратная иерархия резонансов (обратите внимание на логарифмический масштаб графика) и сужение резонансных линий по мере увеличения $n$ (соответствует увеличению резонансной частоты) --- для $n=3$ и $n=4$ линии настолько узкие, что не могут быть разрешены в масштабе данного графика. Подробности в тексте.}}\label{fig:Q_Drude}
\end{figure}

{Приведем пример зависимости полного сечения рассеяния $\sigma_{\rm sca}(\omega) = \sum_n \sigma^{(n)}_{\rm sca}(\omega)$ при фиксированном $R$, рассчитанной на основании точного решения Ми для дисперсионного соотношения, описываемого формулой Друде:

\begin{equation}\label{eq:epsilon_Drude}
  \varepsilon = 1-\frac{\omega_p^2}{\omega^2},\;\; \omega_p = const
\end{equation}
В этом случае условие $\varepsilon = -2$, соответствующее дипольному резонансу ($n=1$) при $r \rightarrow 0$, выполняется при $\omega = \omega_{10} = \omega_p/\sqrt{3}$. Удобно ввести безразмерную эффективность рассеяния \mbox{$Q_{\rm sca} = \sigma_{\rm sca}/(\pi R^2)$,} а $\omega$ измерять в единицах $\omega_{10}$. Что касается параметра размера $x$, то его можно записать в виде  $x=x_{10}\omega/\omega_{10}$, так что \mbox{$x=x_{10}$} при $\omega = \omega_{10}$. Величину же $x_{10}$ для определенности выберем равной 0,3. Результаты соответствующих расчетов приведены на Рис.~\ref{fig:Q_Drude}.    }

{Отметим, что подстановка дисперсионной зависимости \eqref{eq:epsilon_Drude} в рэлеевскую формулу \eqref{eq:sigma_Ray} приводит к выражению
\begin{equation}\label{eq:sigma_Ray_Drude}
  Q_{\rm sca} = \frac{8x_{10}^4(\omega/\omega_{10})^4}{3\left[(\omega/\omega_{10})^2-1\right]^2}.
\end{equation}
В соответствии со сказанным ранее, оно имеет полюс при $\omega=\omega_{10}$. Если же учесть сдвиг резонансной частоты за счет конечности размера частицы и воспользоваться формулой \eqref{eq:an_x<<1_de<<gamma}, то при выбранном значении $x_{10}=0,3$ полюс сдвигается в точку $\omega \approx 0,967\omega_{10}$, которому соответствует $\varepsilon^{\rm res} \approx -2.208$, см. Рис.~\ref{fig:Q_Drude}.     }

\begin{figure}
  \centering
  \includegraphics[width=.85\columnwidth]{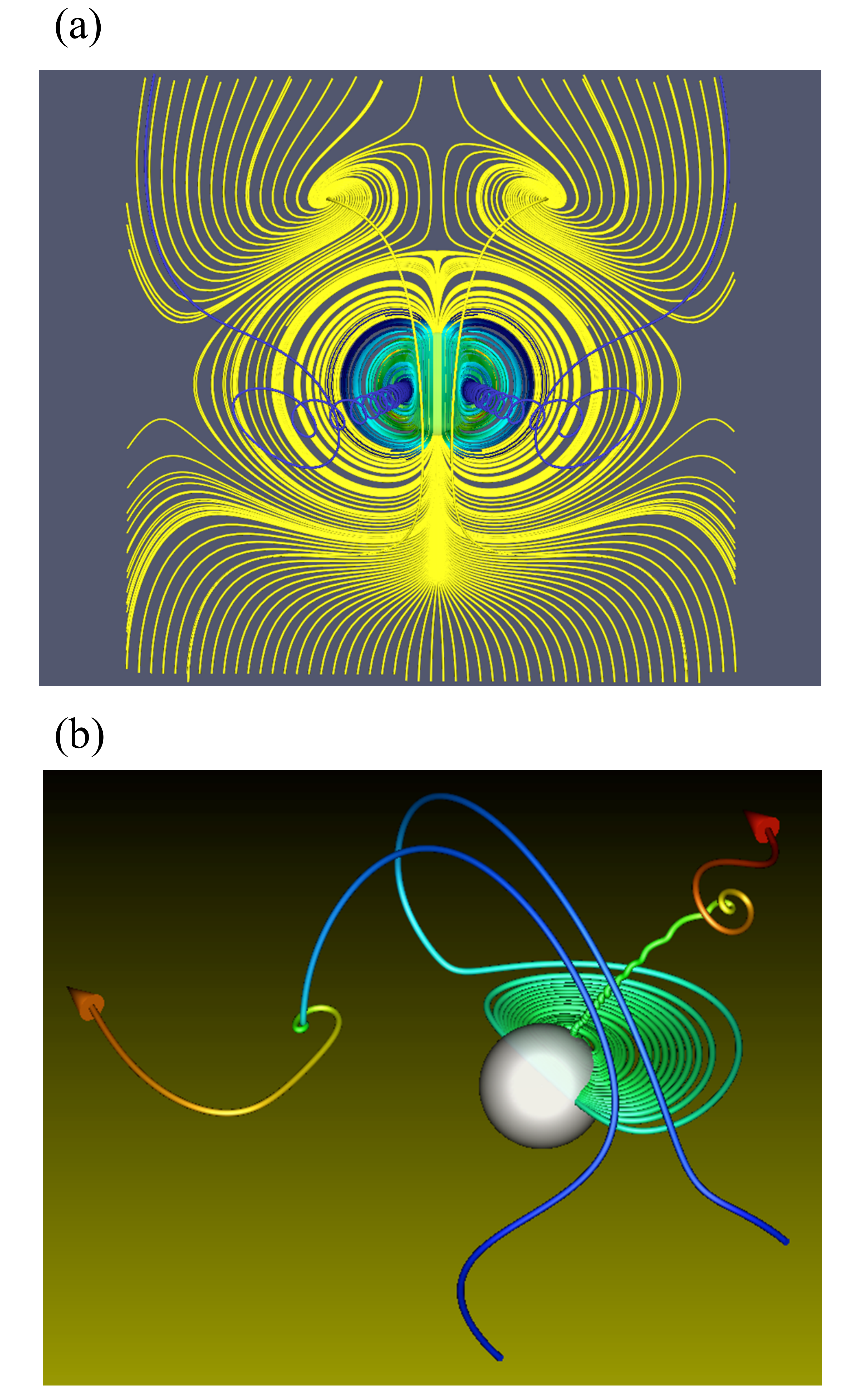}
  \caption{<<Линии тока>> вектора Пойнтинга в окрестности дипольного резонанса; $x=0.3,\; \varepsilon =-2.17$. При данном $x$ точке резонанса соответствует $\varepsilon=-2.22$. (а) --- Падающая плоская волна распространяется от нижнего края панели к верхнему. Вектор $\mathbf{E}$ лежит в плоскости рисунка. В верхней полуплоскости хорошо видны два оптических вихря --- особые точки типа фокус-седло. (b) --- Две отдельные линии тока в 3D. Рассеивающая сфера показана как серый шар. Подчеркнем, что при выбранном значении $x$ длина волны падающего излучения более чем в двадцать раз превышает радиус сферы, так что вся сложная структура поля, приведенная на рисунке, имеет существенно субволновые размеры.}\label{fig:3D}
\end{figure}

Но, пожалуй, самой интересной характеристикой резонансного рассеяния является топологическая структура поля вектора Пойнтинга, определяющая циркуляцию энергии в окрестности рассеивающей сферы. Поскольку размер сферы мал по сравнению с длиной волны излучения, характерным масштабом, определяющим структуру поля в окрестности частицы, является единственный, имеющийся в нашем распоряжении масштаб, --- радиус сферы $R$. Однако в пределах этого масштаба поле имеет сложную структуру, содержащую особые точки, число, тип и пространственное положение которых чрезвычайно чувствительны к малым изменениям частоты падающего излучения, и связанными с ним изменениями $\varepsilon$~\mbox{\cite{Luk2004:PRB,Zheludev2005:OE,Tribelsky:2006jf,luk2008we,tribelsky2014:Springer}}.

В качестве примера на Рис.~\ref{fig:3D} представлены <<линии тока>> поля вектора Пойнтинга в окрестности дипольного резонанса. В каждой точке линий тока вектор Пойнтинга направлен к ним по касательной. Видно, что в структуре поля образуется <<воронка>>, которая втягивает поле из большой области пространства, находящегося <<ниже по течению>> и транспортирует его в малую пространственную область, занятую локализованными плазмонами. Эффект воронки и обеспечивает высокую концентрацию электромагнитной энергии в частице, необходимую для поддержания высокой интенсивности рассеиваемого излучения, которое <<выстреливается>> из центра плазмонов в направлении перпендикулярном плоскости поляризации падающего излучения (нанопрожектор!). Это особенно хорошо видно на Рис.~\ref{fig:3D}(b).

Столь большое число качественных отличий такого резонансного рассеяния от рэлеевского явилось основанием для присвоения ему термина \emph{аномальное рассеяние}~\cite{Libenson2005}. Для удобства читателя эти различия сведены в Таблицу~\ref{tab:Anomal_features}. При этом подразумевается, что параметры задачи точно соответствуют центру соответствующей резонансной линии.
\vspace*{32pt}

\begin{table}[h]
\caption{\label{tab:Anomal_features}%
Сравнительные характеристики рэлеевского и аномального рассеяния
}
\mbox{}\\
\begin{ruledtabular}
\begin{tabular}{lll}
Характеристика                                                                     & \begin{tabular}[c]{@{}l@{}}Рэлеевское\\ рассеяние\end{tabular}                                                                                    & \begin{tabular}[c]{@{}l@{}}Аномальное\\ рассеяние\end{tabular}                                                                                    \\ \hline
\begin{tabular}[c]{@{}l@{}}Иерархия\\ резонансов\end{tabular}                      & \begin{tabular}[c]{@{}l@{}}Нормальная\\ (амплитуды \\ убывают с\\ ростом $n$
\end{tabular} & \begin{tabular}[c]{@{}l@{}}Обратная\\ (амплитуды \\ возрастают с\\ ростом $n$)
\end{tabular}
\\ \hline
\begin{tabular}[c]{@{}l@{}}Зависимость \\ от $R$\end{tabular}                      & $\sigma_{\rm sca}^{(a)}\sim R^{2(2n+1)}$                                                                                               & \begin{tabular}[c]{@{}l@{}}$\sigma_{\rm sca}^{(a)}$\\ не зависит от $R$\end{tabular}                                                   \\ \hline
\begin{tabular}[c]{@{}l@{}}Частотная \\ зависимость\end{tabular}                   & непрерывная

                                                                                   & дискретная
\\ \hline
\begin{tabular}[c]{@{}l@{}}Амплитуды по-\\лей рассеянной\\ волны в окрест-\\ ности сферы\end{tabular} & \begin{tabular}[c]{@{}l@{}}$\mathbf{E}$ и $\mathbf{H}$ регулярны\\ по $R$ при $R \rightarrow 0$\end{tabular}                                         & \begin{tabular}[c]{@{}l@{}}$\mathbf{E}$ и $\mathbf{H}$ сингулярны\\по $R$ при $R \rightarrow 0$\end{tabular}                                                          \\ \hline
\begin{tabular}[c]{@{}l@{}}Поле вектора\\ Пойнтинга\end{tabular}                   & \begin{tabular}[c]{@{}l@{}}Имеет простую\\ структуру. Может \\ содержать\\ сингулярности \\ типа седел \cite{Luk2004:PRB}\end{tabular}                               & \begin{tabular}[c]{@{}l@{}}Имеет сложную\\ структуру. Может \\ содержать \\ различные сингу- \\ лярности, включая\\ оптические вихри\end{tabular} \\ 
\end{tabular}
\end{ruledtabular}
\end{table}

\subsubsection{Эффект конечной диссипации}

Обсудим теперь случай конечной, но малой диссипации. При наличии диссипативных потерь диэлектрическая проницаемость становится комплексной величиной: \mbox{$\varepsilon=\varepsilon'+i\varepsilon''$}. Если диссипация мала \mbox{($0<\varepsilon''\ll 1$)}, то ее учет сводится к тривиальной замене в выражении \eqref{eq:an&gamma_n_x<<1} для $a_n$ действительного $\delta\varepsilon$ на \mbox{$\delta\varepsilon'+i\varepsilon''$}. Это приводит к увеличению ширины линии, что проявляется в замене в знаменателе $\gamma_n$ на \mbox{$\gamma_n + 2\varepsilon''$}, где первый член описывает недиссипативное радиационное затухание, а второй --- обычные диссипативные потери.

Очевидно, что аномальное рассеяние может реализовываться только при условии преобладания радиационного затухания над диссипативным, т.е. при  \mbox{$\gamma_n \gg \varepsilon''$}. В обратном предельном случае радиационным затуханием можно пренебречь. Легко видеть, что это эквивалентно пренебрежению $F$ по сравнению с $G$ в \eqref{eq:an_FG}, т.е. переходу к рэлеевскому приближению. Переход от одного режима к другому происходит при~\cite{Tribelsky:JETP:1984}
\begin{equation}\label{eq:crossover}
  \varepsilon'' \sim \frac{(n+1)x^{2n+1}}{[n(2n-1)!!]^2}.
\end{equation}

Отсюда становится ясным, что нет необходимости обобщать наши рассуждения на случай немалых значений $\varepsilon''$ --- при переходе к немалой диссипации восстанавливается рэлеевское рассеяние, и вопрос снимается сам по себе.

Выражение \eqref{eq:crossover} также разрешает и отмечавшийся выше парадокс о конечном сечении аномального рассеяния для частицы нулевого размера в недиссипативном пределе. Недиссипативный предел --- это идеализация, которая никогда не реализуется в природе. Полного отсутствия диссипативных процессов не бывает. В соответствии с этим, в реальных случаях мнимая часть диэлектрической проницаемости может быть очень мала, но она всегда конечна. С другой стороны, при любом сколь угодно малом, но фиксированном значении $\varepsilon''$ по мере уменьшения $x$ происходит переход от аномального к рэлеевскому рассеянию, и при \mbox{$x \rightarrow 0$} сечение рассеяния обращается в ноль, в полном соответствии с формулой \eqref{eq:sigma_Ray} и здравым смыслом.

Еще один независимый механизм связан с резким сужением ширины резонансной линии при  \mbox{$x \rightarrow 0$}, см. \eqref{eq:an&gamma_n_x<<1}. Поскольку любой реальный источник излучения имеет конечную ширину линии, $\gamma_n$ при достаточно малом значении $x$ неизбежно становится меньше этой ширины, так что только часть излучения поглощается резонансно. При \mbox{$x \rightarrow 0$} эта часть стремится к нулю, что также приводит к обращению в ноль соответствующего сечения~\cite{Brynkin2019}.
\pagebreak

Отметим, что до сих пор мы говорили о пространственно-однородной частице. В последнее время большое внимание уделяется двухслойным частицам типа ядро-оболочка (core-shell) и их более многослойным обобщениям --- оболочечным частицам, состоящим из нескольких слоев с разными оптическими свойствами~\cite{chen2017vortex,shore2015scattering,Garg_2017,hasegawa2006enhanced,li2015atomically,Alu2014cloaked_sensors}. При концентрическом расположении слоев задача о рассеянии плоской монохроматической линейно поляризованной электромагнитной волны такой частицей тоже допускает точное решение.  тем же методом мультипольного разложения, Достаточно подробное обсуждение данного решения можно найти, например, в работе~\cite{shore2015scattering}.

В этом решении рассеянное поле вне частицы представляется в виде того же самого мультипольного разложения, что и в случае однородной сферы, поэтому выражения \eqref{eq:sigma_sph_n}, \eqref{eq:sigma_ext_sph_n}, связывающие различные парциальные сечения с коэффициентами рассеяния, остаются неизменными. Что же касается самих коэффициентов рассеяния, то хотя они по-прежнему сохраняют ту же фундаментальную структуру, определяемую формулой~\eqref{eq:an_FG}, конкретные выражения для функций $F_n,\; G_n$ становятся тем более сложными, чем больше концентрических слоев имеет частица (в общем случае произвольного числа слоев они определяются через некоторое число рекуррентных соотношений). При этом задача приобретает важное качественно новое свойство: меняя радиусы ядра и слоев, а также их диэлектрические проницаемости, можно сдвигать положения различных резонансов и добиваться того, что несколько различных резонансов могут возбуждаться на одной и той же частоте (вырождение резонансов).

В силу линейности задачи и ортогональности собственных мод, парциальные сечения различных резонансов складываются. Каждый из резонансов может быть оптимизирован с точки зрения максимизации соответствующего парциального сечения рассеяния. Совмещая требования оптимизации с условиями вырождения определенных резонансов, можно сконструировать суперрасеивающую оболочечную структуру, суммарное сечение рассеяния которой может значительно превосходить максимальное значение этой величины для одной парциальной моды, подробнее см.~\cite{Fan:PRL:2010}.

Завершая обсуждение теоретического описания аномального рассеяния рассмотрим, какие отличия возникают при рассеянии света бесконечным цилиндром кругового сечения. В основном, все сказанное о сфере переносится на случай цилиндра без существенных изменений~\cite{LukTribTer:JOptTech2006,luk2007peculiaritiesJOA,luk2007peculiaritiesCOLA}. Однако между этими двумя случаями имеется одно качественное отличие: если для сферы при \mbox{$R\rightarrow 0$} резонансные значения $\varepsilon$ для различных мультиполей различны и перекрытия резонансов не происходит, то для цилиндра в том же пределе резонансное  $\varepsilon$ для всех $n$ кроме \mbox{$n=0$} одно и то же и равно -1. В таком случае возникает вопрос о сходимости мультипольного разложения при \mbox{$\varepsilon\rightarrow-1,\;R\rightarrow 0$}. Этот вопрос подробно изучен в работе~\cite{Brynkin2019}. Показано, что для коэффициентов рассеяния точка \mbox{$\varepsilon\rightarrow-1,\;R\rightarrow 0$} является аналогом существенно особых точек аналитических функций, так что коэффициенты рассеяния не имеют в этой точке определенного предела (см., также МД). Для устранения возникающей в связи с этим проблемы, как только что отмечалось выше, требуется выйти за рамки монохроматического приближения и учесть конечность ширины линии источника, после чего в правильной последовательности осуществить ряд предельных переходов. В результате получается вполне определенное выражение для сечения рассеяния, которое обращается в ноль при \mbox{$R\rightarrow 0$}.

Укажем также, что проблема аномального рассеяния цилиндром имеет давнюю и неожиданную для читателя-оптика историю в связи с радиозондированием метеорных треков в атмосфере~\cite{herlofson1951plasma,kaiser1952theory,closs1953xxxiv}. Пролетая с космическими скоростями через земную атмосферу, метеоры создают почти идеальный цилиндрический трек ионизированного воздуха, который в ряде случаев можно рассматривать как бесстолкновительную плазму. Резонансные частоты такого цилиндра лежат в радиодиапазоне. При зондировании на резонансной частоте трек дает мощный отраженный сигнал, который полностью соответствует рассмотренному здесь аномальному рассеянию. Изучая этот сигнал можно получить подробную информацию об исходном метеоре. Детальное обсуждение этих вопросов лежит вне рамок данного обзора.

\section{Максимизация поглощение света субволновой частицей. Аномальное поглощение}

Другой важный круг вопросов связан с энерговыделением, происходящем в субволновой частице за счет диссипативных потерь при ее облучении электромагнитным излучением. В особенности нас будет интересовать задача максимизации энерговыделения~\cite{Baffou:N:2010}, важная для ряда приложений, например, в медицине --- лечение онкологических заболеваний, когда лазерный нагрев наночастиц, внедренных в опухолевую ткань, приводит к уничтожению раковых клеток на клеточном уровне~\cite{Skirtach:NL:2005,Han:N:2007,Brigger:ADDR:2002,Huang:LMS:2007}, а также в задачах микрохирургии~\cite{Anderson:S:1983}; в проблеме высоко-плотной записи информации~\cite{Pan:NP:2009} (нагрев наночастицы выше температуры Кюри приводит к локальному размагничиванию магнитных материалов) и др.

Разумеется, коль скоро речь заходит о приложениях, то само по себе энерговыделение интереса не представляет. Важно знать, не сколько энергии выделилось в частице, а до какой температуры эта частица нагрелась. Ответ на этот вопрос дает совместное решение уравнений Максвелла и уравнений теплопереноса как в самой частице, так и в окружающей ее среде. Однако в линейной задаче эти две проблемы разделяются --- решение электромагнитной части задачи входит в задачу теплопереноса только в виде источника и не зависит от решения последней. Поскольку задачи теплопереноса лежат вне рамок данного обзора, здесь мы ограничимся только изучением энерговыделения. Что же касается полной задачи, включающей теплоперенос, то ее обсуждение можно найти, например, в  работах~\cite{luk2012paradoxesNJP,tribelsky2011:PRX,Tribel_Fukum2016_BOE}.

На первый взгляд, максимизация поглощения не вызывает никаких проблем. Для рассматриваемых немагнитных материалов плотность энерговыделения за счет диссипативных потерь пропорциональна $\varepsilon''|\mathbf{E}|^2$~\cite{LL_8} --- чем больше мнимая часть проницаемости, тем больше поглощение.

В действительности же все несколько сложнее:  $\mathbf{E}$, которое входит в это выражение, это поле в той точке пространства, где вычисляется плотность энерговыделения. Энерговыделение происходит внутри частицы, следовательно $\mathbf{E}$ --- это поле \emph{внутри} частицы.

При резонансе поле внутри частицы возрастает. Если рассматривать возбуждение резонансных мод внешним полем как вынужденные колебания гармонического осциллятора, то в стационарном состоянии амплитуда колебаний в точке резонанса оказывается обратно пропорциональной диссипативной константе. В нашем случае это означает что в окрестности резонанса поле внутри частицы по порядку величины оценивается выражением $\mathbf{E_0}/\varepsilon''$, где $\mathbf{E_0}$ --- амплитуда поля падающей волны. В результате плотность энерговыделения оказывается пропорциональной $|\mathbf{E_0}|^2/\varepsilon''$.

Теперь мы приходим к прямо противоположному выводу: для максимизации энерговыделения мы должны использовать в качестве материала сферы вещество с малыми диссипативными потерями и при этом подбирать частоту излучения так, чтобы она оказалась в окрестности резонансной частоты.

Вспомним теперь результаты предыдущего раздела --- при уменьшении $\varepsilon''$ рассеяние в окрестности резонансов неизбежно становиться аномальным, см. \eqref{eq:crossover}. Поэтому для корректного рассмотрения проблемы максимизации поглощения опять необходимо рассмотреть точное решение задачи.

Это несложно сделать. Согласно точному решению, парциальное сечение поглощения для сферы описывается следующим выражением~\cite{Kerker2013,Bohren::1998}:
\begin{equation}\label{eq:sigma_sph_abs_n}
  \sigma_{\rm abs}^{(n)}  = \frac{{2(2n + 1)\pi}}{{k^2}}\left[\left({\rm Re}\,a_n-|a_n  |^2\right)  +\left({\rm Re}\,b_n-|b_n  |^2\right)\right],
\end{equation}

Имея в виду более широкое применение результатов, не будем пока ограничиваться предположением о малости $x$. Поскольку электрические и магнитные моды независимы, задача максимизации $\sigma_{\rm abs}^{(n)}$ сводится к поиску максимума выражения \mbox{$z_n^\prime - z_n^{\prime\,2} - z_n^{\prime\prime\,2}$}, где вместо $z_n$ может фигурировать как $a_n$, так и $b_n$ Легко видеть, что максимуму этого выражения соответствует \mbox{$z_n^{\prime\prime}=0,\; z_n^\prime=1/2$}, а сам максимум равен 1/4. При этом значение парциального сечения поглощения оказывается равным парциальному сечению рассеяния и равным $\sigma_{\rm sca\,max}^{(n)}/4$.

Отметим, что этот результат эквивалентен хорошо известному в теории антенн критерию максимизации мощности, передаваемой от источника излучения к его приемнику~\cite{kraus2002antennas,balanis2016antenna}. {Из него, в частности, следует, что эффективная площадь малой резонансной антенны (сечение поглощения) оказывается порядка квадрата длины волны излучения и не зависит от размеров антенны --- иначе карманный транзисторный приемник не мог бы принимать радиопередачи в метровом диапазоне (!), ср. с обсуждавшимся выше результатом для аномального поглощения: $\sigma_{abs} \sim 1/k^2$ }. Более общие результаты, справедливые для рассеяния и поглощения света частицей произвольной формы, содержатся в работах\mbox{~\cite{PhysRevLett.120.033902_2018-UltimateAbsorption,miller2016fundamental,Ruan2012}}.

Полезно также подчеркнуть, что, хотя в общем случае коэффициенты рассеяния комплексны, в максимуме аномального рассеяния (при нулевой диссипации) и в максимуме поглощения (при конечной диссипации) они становятся чисто действительными величинами, равными 1 и 1/2, соответственно. Это означает, что в этих случаях поля рассеянной и падающей волны оказываются синфазны.

Заметим, что при комплексном $\varepsilon$ функции $F$ и $G$ в выражении \eqref{eq:an_FG} также становятся комплексными. Записывая их в виде \mbox{$F=F'+iF''$}, \mbox{$G=G'+iG''$} перепишем выражения для амплитуд рассеяния в виде
\begin{equation}\label{eq:z_n_FG_complex}
  z_n = \frac{F'_n+iF''_n}{F'_n-G''_n+i(F''_n+G'_n)}.
\end{equation}
В этом случае после несложных вычислений условия максимума поглощения \mbox($z'_n=1/2,\;z''_n=0$) переписываются в виде \emph{соотношений взаимности}: \mbox{$F'_n=-G''_n,\;F''_n=G'_n$}.

Воспользуемся теперь малостью $x$. Как уже отмечалось, в этом случае при любых $n$ для магнитных мод функции $G_n$ порядка единицы и нигде не обращаются в ноль, а функции $F_n$ малы. В результате поглощение, связанное с возбуждением магнитных мод, тоже мало и не представляет для нас интереса.

Что же касается электрических мод, то, в соответствии с \eqref{eq:F_an_x<<1}, \eqref{eq:G_an_x<<1},  при \mbox{$\varepsilon'' \ll 1$} в главном приближении $F_n$ можно считать чисто действительной величиной. Тогда, в соотношениях взаимности можно положить \mbox{$F''_n=0$}, откуда немедленно следует, что  \mbox{$G'_n=0$}. Это условие в точности совпадает с условием максимизации аномального рассеяния при отсутствии диссипации. Второе же условие взаимности \mbox{$F'_n=-G''_n$} с учетом \eqref{eq:F_an_x<<1}, \eqref{eq:G_an_x<<1} приводит к тому, что максимум поглощения реализуется точно посередине переходной области от аномального рассеяния к рэлеевскому, когда в формуле \eqref{eq:crossover} знак $\sim$ заменяется на знак равенства, и вклады, вносимые в ширину линии радиационным и диссипативным затуханием оказываются равны друг другу.

Таким образом, для субволновой частицы максимум парциального сечения поглощения связан с резонансным возбуждением\emph{ электрических} мод и достигается при \emph{малом\/} значении мнимой части диэлектрической проницаемости. При это само рассеяние оказывается близко к аномальному и в значительной степени наследует все его особенности, см. Таблицу~\ref{tab:Anomal_features}. Такое поглощение также естественно назвать аномальным~\cite{Tribelsky:2011ir}.

Что можно сказать о форме линий при аномальном поглощении? Прежде всего, отметим, что при острых резонансах ширина линии определяется поведением знаменателя выражения \eqref{eq:an_FG}. Знаменатели у парциальных сечений рассеяния и поглощения одинаковы, см. \eqref{eq:sigma_sca_n_max}, \eqref{eq:sigma_sph_abs_n}. Поэтому обе линии имеют одинаковую форму и отличаются только масштабным множителем. Если, как это принято, под формой линии понимать зависимость соответствующего парциального сечения от $\omega$, то ничего необычного здесь нет. Фактически, этот вопрос уже обсуждался в предыдущем разделе, где было показано, что линия имеет колоколообразную лоренцевскую форму, и была определена ее ширина, в том числе и с учетом диссипации. Результаты эти полностью переносятся и на настоящий случай.

Однако имеет смысл расширить стандартное определение формы линии и рассмотреть $\sigma^{(n)}_{\rm abs}$, как функцию \emph{двух\/} независимых переменных: $\varepsilon'$ и $\varepsilon''$, или, что то же самое, как функцию комплексного $\varepsilon$. Воспользовавшись тем, что для субволновой частицы в главном приближении $F_n$ --- чисто действительные функции, а \mbox{$G_n \sim \delta\varepsilon \equiv \delta\varepsilon'+i\varepsilon''$},  удобно в окрестности максимума поглощения вместо $\sigma^{{\rm sph}\,(n)}_{\rm abs},\; \varepsilon'$ и $\varepsilon''$ ввести новые безразмерные переменные:
\begin{equation}\label{eq:kappa_xi_zeta}
  \kappa = \frac{x^2 \sigma^{{\rm sph}\,(n)}_{\rm abs}}{2(2n+1)\pi R^2},\;\xi = -\frac{G^{(a)\prime}_n}{F^{(a)\prime}_n},\;\zeta = -\frac{G^{(a)\prime\prime}_n}{F^{(a)\prime}_n}.
\end{equation}
Во избежание недоразумений отметим, что в силу условий взаимности и того, что \mbox{$G''_n \sim \varepsilon'' > 0$} следует, что в максимуме поглощения \mbox{$F^{(a)\prime}_n = -G^{(a)\prime\prime}_n <0$}. По непрерывности те же знаки $F$ и $G$ сохраняются и в окрестности максимума, так что $\zeta >0$.

\begin{figure}
  \centering
  \includegraphics[width=.9\columnwidth]{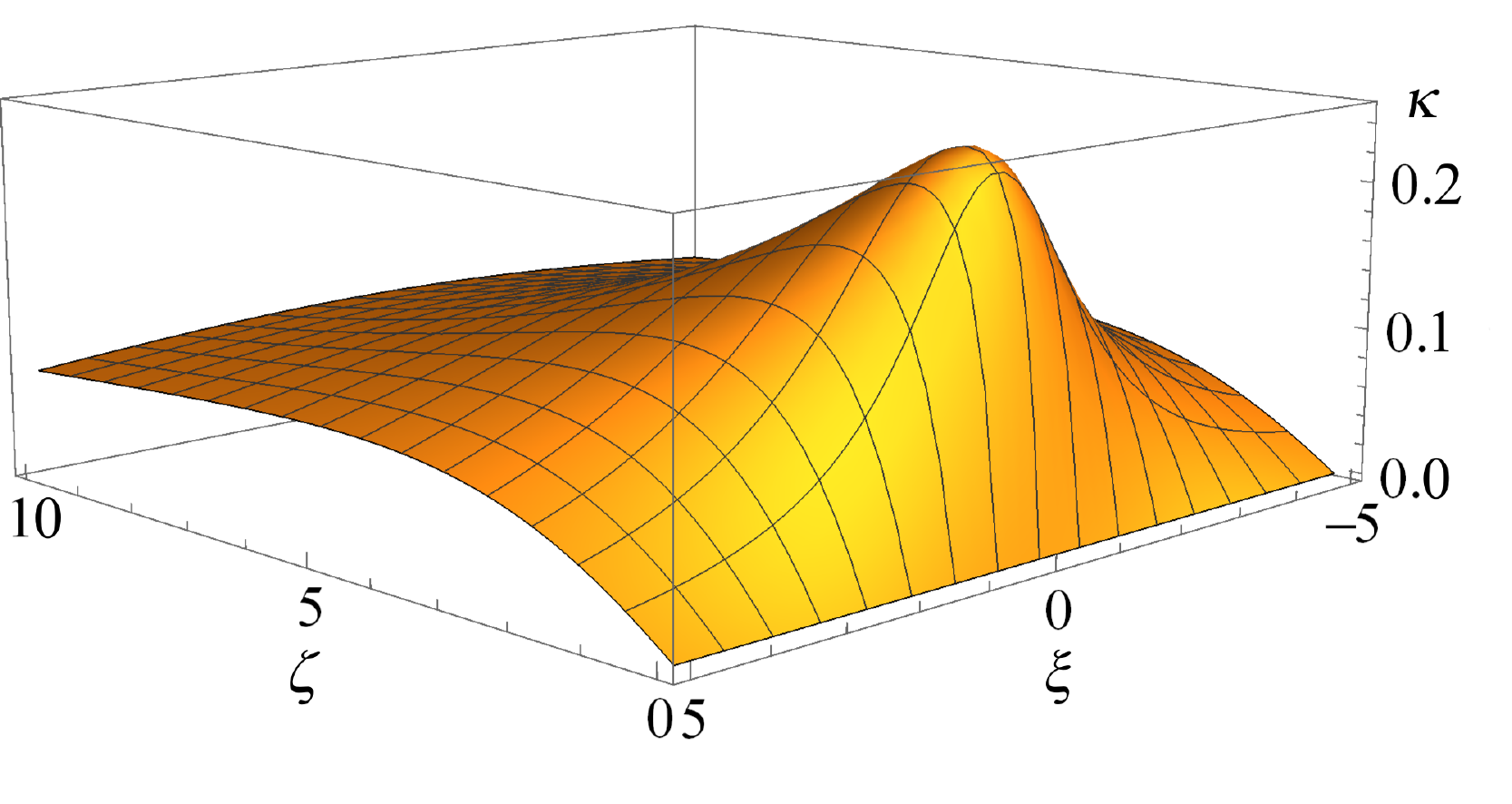}
  \caption{Универсальная поверхность $\kappa=\kappa(\xi,\zeta)$. Подробности в тексте. }\label{fig:kappa}
\end{figure}

Легко видеть, что в этом случае при любом $n$ поверхности $\sigma^{{\rm sph}\,(n)}_{\rm abs}(\varepsilon',\varepsilon'')$ сводятся к единой универсальной поверхности, описываемой уравнением~\cite{Tribelsky:2011ir}
\begin{equation}\label{eq:kappa(xi,zeta)}
  \kappa = \frac{\zeta}{(1+\zeta)^2+\xi^2}, \;\; \zeta >0.
\end{equation}
Эта поверхность изображена на Рис. \ref{fig:kappa}. Отметим, что если сечение поверхности $\kappa=\kappa(\zeta,\xi)$ плоскостями \mbox{$\zeta=const$} дает симметричный лоренцевский профиль, то сечения плоскостями  $\xi=const$ порождают семейство сильно несимметричных кривых с резким спадом при изменении $\zeta$ от точки, соответствующей максимуму профиля, в направлении уменьшения $\zeta$ и длинным, медленно спадающим хвостом при изменении $\zeta$  в противоположном направлении, см. также~\cite{tretyakov2014maximizing}.

\section{Аномальное поглощение и реальные материалы}

Рассуждения, приведенные в предыдущих разделах, молчаливо предполагали, что самое главное --- определить надлежащие условия для реализации обсуждаемых там эффектов. После того, как это сделано, надо просто выбрать нужный образец для наилучшего наблюдения этих эффектов. Однако беда в том, что, если оставить в стороне искусственные метаматериалы, специально изготовленные для того, чтобы иметь заданные свойства, и ограничиться обычными, существующими в природе, то они имеют вполне определнные дисперсионные зависимости $\varepsilon(\omega)$. Поэтому задача состоит не в том, чтобы <<выбрать нужный образец>>, а в том чтобы понять, существует ли вообще материал с необходимой зависимостью $\varepsilon(\omega)$. {Проблема эта подробно обсуждалась в наших работах~\cite{Tribelsky:2011ir,PhysRevLett.120.033902_2018-UltimateAbsorption}, но теперь принятая там манера обсуждения представляется нам неудачной, и ниже мы применим другой, эквивалентный подход.}

{Как показано выше, условия аномального поглощения предполагают, что в этом случае Re$\,a_n = 1/2$, Im$\,a_n=0$. С другой стороны, коэффициенты рассеяния есть определенные функции параметров задачи, однозначно заданные точным решением Ми, согласно которому $a_n=a_n(x,mx)$, см. МД. Будем для простоты считать, что рассеивающая излучения частица находится в вакууме, так что $m_{\rm out} \equiv 1$. Что же касается $m=\sqrt{\varepsilon}$, то для данного материала частицы эта величина однозначно определяется его дисперсионной зависимостью, которая в случае наночастицы может дополняться еще и зависимостью от размера частицы $R$. В результате, условия аномального поглощения приобретают вид
\begin{equation}
\begin{aligned}\label{eq:Re_an=1/2_Im=0_matter}
  & {\rm Re}\left[a_n\left(\frac{R\omega}{c},\varepsilon(\omega,R)\right)\right]  = \frac{1}{2},\\
  & {\rm Im}\left[a_n\left(\frac{R\omega}{c},\varepsilon(\omega,R)\right)\right]  = 0.
\end{aligned}
\end{equation}
Эти условия, рассматриваемые как уравнения, должны определять дискретные пары \mbox{$\omega$ и $R$} для которых реализуется аномальное поглощение.} {Увы, до сих пор не удалось найти \emph{ни одного\/} реального вещества, для которого эти уравнения имели бы решение в оптическом диапазоне частот. Означает ли это, что аномальное поглощение остается гипотетическим эффектом, который никогда нельзя наблюдать в реальном эксперименте? Не вполне. Если абсолютная истина недостижима, то к ней, по крайней мере, можно приблизиться! Если уравнения \eqref{eq:Re_an=1/2_Im=0_matter} не могут быть выполнены точно, то надо попытаться найти их приближенное решение.}

\begin{figure}
  \centering
  \includegraphics[width=.95\columnwidth]{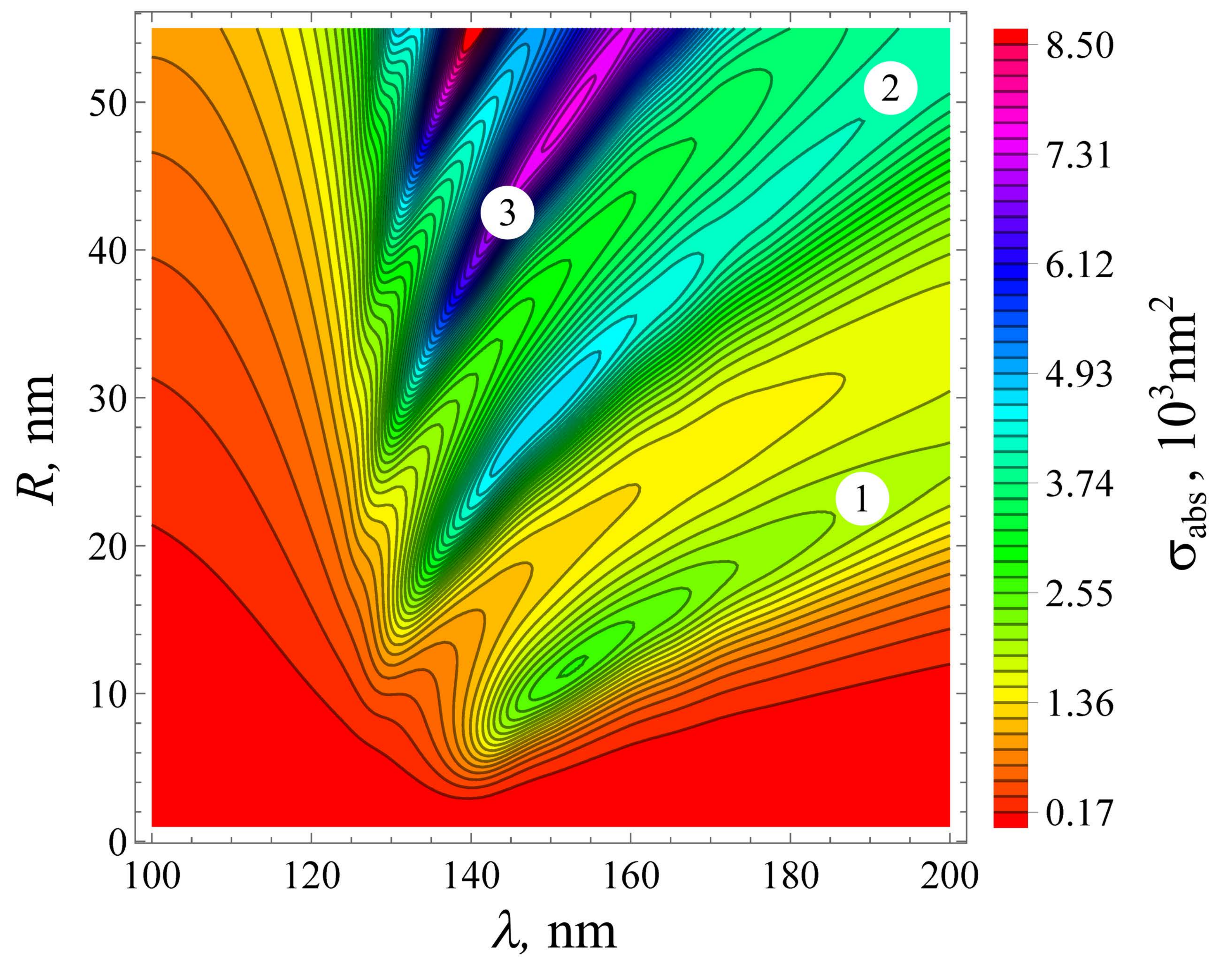}
  \caption{Сечение поглощения алюминиевой сферы как функция ее радиуса и длины волны излучения. Расчет по формулам Ми с учетом реальных оптических свойств наночастицы алюминия~\cite{Tribelsky:2011ir}, см. подробнее в тексте. Области с преимущественным вкладом дипольного, квадрупольного и октупольного моментов отмечены цифрами 1, 2 и 3, соответственно. Хорошо видна обратная иерархия резонансов.}\label{fig:Al}
\end{figure}

Предыдущий анализ показывает, что для субволной частицы это решение нужно искать для таких материалов, которые в оптическом диапазоне имеют $\varepsilon' \approx -2$ и малое $\varepsilon''$. Такими свойствами обладает, например, алюминий. В соответствии с этим, в работе~\cite{Tribelsky:2011ir} были проведены расчеты сечения поглощения алюминиевой сферы при ее облучении в вакууме плоской, линейно поляризованной электромагнитной волной с циклической частотой $\omega$. Расчеты проводились  на основе точного решения Ми с учетом реальной экспериментальной зависимости диэлектрической проницаемости от частоты, в которых также феноменологически учитывалась ее зависимость от радиуса сферы $R$ за счет изменения частоты столкновений электронов при $2R<\tau v_{_F}$, где $v_{_F}$ --- фермиевская скорость, а $\tau$ --- частота электронных столкновений. Результаты расчетов представлены на Рис.~\ref{fig:Al}, где хорошо видна обратная иерархия резонансов.

Согласно данным, приведенным на Рис.~\ref{fig:Al}, максимум дипольного поглощения достигается при \mbox{$R \approx 11,8$} нм на длине волны \mbox{$\lambda \approx 152,8$} нм. При этом сечение поглощения оказывается равным \mbox{$2,74\cdot10^3$ нм$^2$}. Расчет же по формуле $\sigma_{\rm sca\,max}^{(n)}/4$, определяющий абсолютный максимум парциального сечения при аномальном поглощении, дает \mbox{$ 2,79\cdot10^3$ нм$^2$} --- различие между этими величинами только в третьем знаке. Отметим также, что значение геометрического экваториального сечения сферы $\pi R^2$ при данном значении $R$ составляет \mbox{$ 0,44\cdot10^3$ нм$^2$}, т.е. величину в 6,22 раза меньшую ее сечения поглощения, что является убедительным доказательством эффекта воронки~\cite{bohren1983can}.

{Обсудим теперь, что нового вносят в проблему многослойные частицы, уже упоминавшиеся выше в связи с проблемой аномального рассеяния. Для простоты ограничимся простейшим случаем частицы типа ядро-оболочка, где вокруг сферического ядра радиуса $R_1$ располагается концентрическая оболочка с внешним радиусом $R_2$, см. врезку на Рис.~\ref{fig:core-shell}. В этом случае условия аномального поглощения по-прежнему гласят, что \mbox{Re$\,a_n=1/2$}, \mbox{Im$\,a_n=0$}. Однако коэффициенты рассеяния теперь зависят от б\'{о}льшего числа параметров. Это дает нам дополнительные степени свободы, конкретная реализация которых зависит от вида дисперсионных зависимостей диэлектрической проницаемости ядра и оболочки.}

{В качестве примера рассмотрим случай ядра, сделанного из материала с сильной дисперсионной зависимостью и непоглощающей излучение оболочки, для которой зависимостью $\varepsilon$ от параметров задачи можно пренебречь. Тогда вместо \eqref{eq:Re_an=1/2_Im=0_matter} мы получаем следующие уравнения:}
{
\begin{equation}
\begin{aligned}\label{eq:Re_an=1/2_Im=0_shell}
  & {\rm Re}\left[a_n\left(\frac{R_1\omega}{c},\frac{R_2\omega}{c},\varepsilon_1(\omega,R_1,R_2),\varepsilon_2\right)\right]  = \frac{1}{2},\\
 & {\rm Im}\left[a_n\left(\frac{R_1\omega}{c},\frac{R_2\omega}{c},\varepsilon_1(\omega,R_1,R_2),\varepsilon_2\right)\right]  = 0.
\end{aligned}
\end{equation}
где индекс 1 относится к ядру, 2 --- к оболочке и $\varepsilon_2=const$ (Im$\,\varepsilon_2=0$). Теперь мы имеем два условия для четырех параметров: $R_{1,2},\;\varepsilon_2$ и $\omega$. Это позволяет либо подбирать значения параметров задачи для наблюдения аномального поглощения в желаемой области, либо потребовать совпадения резонансных значений параметров одновременно для двух различных значений $n$. При этом частица ведет себя как <<суперпоглощающая>>, т.к. ее полное сечение поглощения, в которое теперь входят два максимальных парциальных, будет существенно превышать абсолютный верхний предел, существующий для каждой из этих парциальных мод в отдельности. Отметим, что такой механизм конструирования суперпоглощающей частицы вполне аналогичен созданию <<суперрассеивающей>> частицы, обсуждавшейся в работе~\cite{Fan:PRL:2010}.

{Важно подчеркнуть, что все поглощение осуществляется в ядре. Диэлектрическая оболочка только заставляет металлическое ядро более эффективно поглощать падающее излучение. Никакого энерговыделения в самой оболочке не происходит.}

\begin{figure}
  \centering
  \includegraphics[width=\columnwidth]{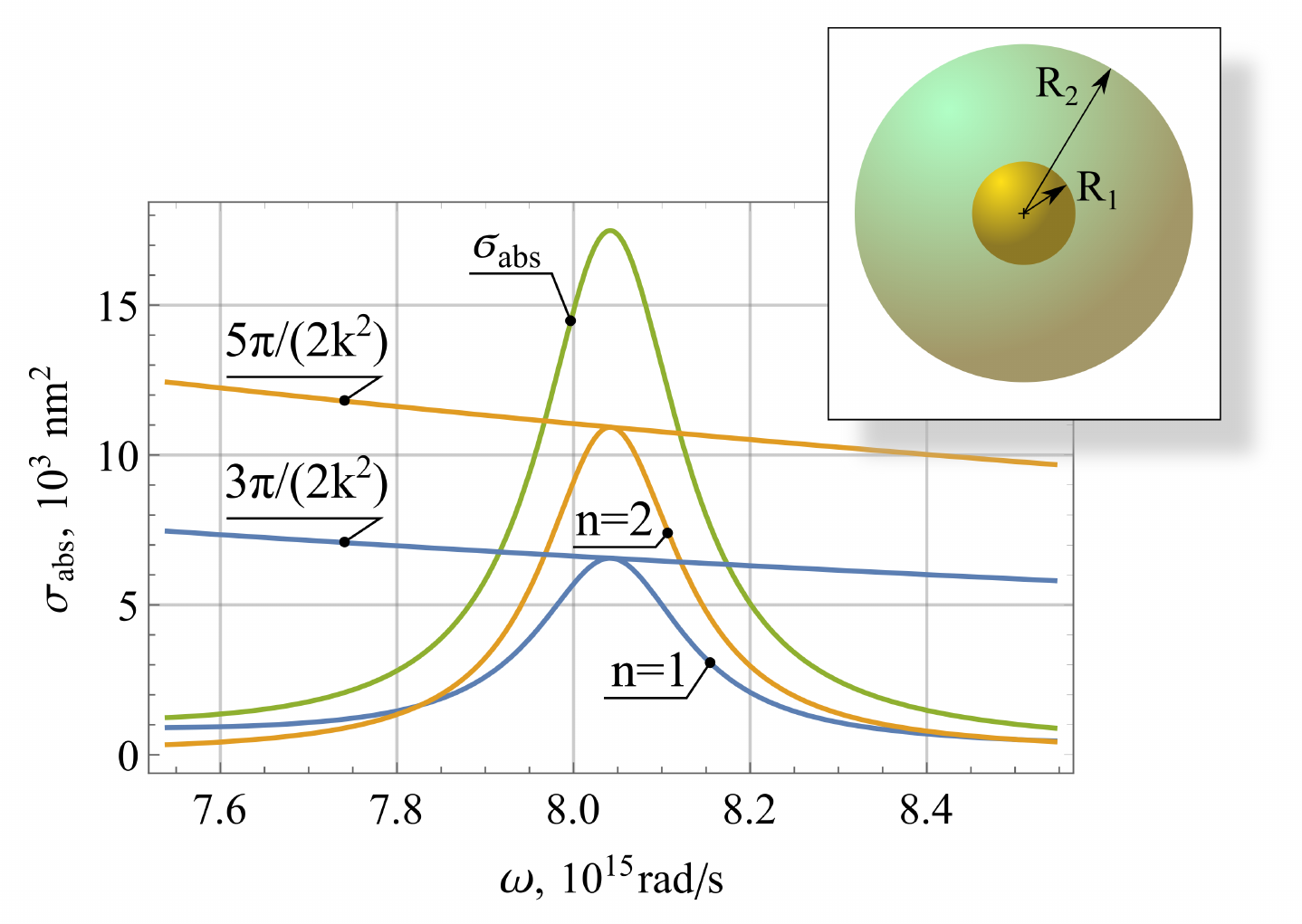}
  \caption{{Двойное резонансное поглощение частицей с золотым ядром и непоглощающей диэлектрической оболочкой (схематически изображена на врезке). Форма линий парциальных сечений поглощения для электрических дипольной ($n=1$) и квадрупольной ($n=2$) мод, а также полного сечения поглощения частицы $\sigma_{\rm abs}$ как функции частоты рассеиваемой волны. Остальные параметры задачи фиксированы и равны их резонансным значениям~\cite{PhysRevLett.120.033902_2018-UltimateAbsorption} (Supplemental Material). Прямые линии соответствуют максимально возможным значениям парциальных сечений поглощения $(2n+1)\pi/(2k^2)$.}}\label{fig:core-shell}
\end{figure}

Конечно, как и в случае с однородной частицей, полученные уравнения не всегда могут быть разрешимы. Однако наличие свободных параметров, делает задачу максимизации поглощения для слоистой частицы значительно более гибкой, чем в пространственно-однородном случае. В частности, вычисления, приведенные в Supplemental Material к работе~\cite{PhysRevLett.120.033902_2018-UltimateAbsorption} {показывают, что для частицы, с металлическим ядром (золото) и недиссипативной диэлектрической оболочкой полное перекрытие линий, связанных с возбуждением идеального аномального поглощения в дипольной и квадрупольной модах, происходит при следующих значениях параметров задачи: $R_1=6,18$ нм, $R_2=31,04$ нм, $\varepsilon_2 = 3,55$ и $\omega = 8,04 \cdot 10^{15}$ рад/с, что соответствуют воздействию на наночастицу излучения УФ диапазона, см. также Рис.~\ref{fig:core-shell}.}

Что касается экспериментального подтверждения изложенных здесь результатов, то, несмотря на очевидную важность проблемы, как с академической точки зрения, так и для большого числа приложений, достоверные эксперименты в оптическом диапазоне, в которых нашли бы подтверждение (или опровержение) обсуждаемые здесь свойства аномального рассеяния и/или поглощения авторам неизвестны. Мы надеемся, что данный обзор может послужить стимулом для проведения таких экспериментов.

\section{Резонансы Фано}
\subsection{История вопроса}

{Перейдем теперь к другому важному типу резонансов: резонансам Фано, названным так в честь } \'{У}го Ф\'{а}но --- ученика Энрико Ферми. Сначала остановимся кратко на их истории, поскольку она и занимательна, и поучительна. По-видимому, впервые резонансы Фано наблюдались в 1902 г. Робертом Вудом при исследовании дифракции света на отражающей дифракционной решетке (аномалии Вуда)~\cite{wood1902remarkable}, а их первое объяснение было дано в 1907 г. лордом Рэлеем~\cite{rayleigh1907dynamical}. При чем здесь Фано? А вот при чем.

В 1935 г. Ганс Бойтлер обнаружил в спектре благородных газов линии поглощения необычной асимметричной формы, которые к тому же наблюдались в области, соответствующей непрерывному спектру, где вообще никаких линий поглощения быть не должно~\cite{Beutler:1935}. Энрико Ферми, который тогда был профессором римского университета Ла Сапиенца, предложил объяснить эти эксперименты своей группе, известной как ``Ребята с улицы Панисперна'', названной так в честь того, что именно на этой улице в Риме находился Институт Физики. Там и сейчас есть отличный кабачок с тем же названием: ``Ragazzi di via Panisperna''. Группа состояла из молодых физиков. Все они впоследствии получили мировую известность. Это были Эмилио Джино Сегре, Бруно Понтекорво, Этторе Майорана и совсем еще молодой, \mbox{23-летний} Уго Фано. Именно он блестяще справился с поставленной Ферми задачей.

Нелишне отметить, что в оставшемся после смерти Ферми архиве было найдено ее полное решение, сделанное им либо перед постановкой задачи для ребят с улицы Панисперна, либо непосредственно после того. Однако, когда Фано закончил вычисления и спросил Ферми, можно ли начинать писать совместную статью, Ферми ответил, что писать можно, но не совместную, а за единоличным авторством Фано: <<Вы сделали все сами, достаточно ограничиться мне стандартной благодарностью>>. Статья Фано~\cite{Fano:NC:1935} была опубликована в том же 1935 г. в итальянском журнале \emph{Nuovo Cimento\/} на итальянском языке. Увы.., она осталась незамеченной. На эту оригинальную статью, практически, никто не ссылается.

Прошло много лет. Судьба занесла обоих ученых в США, где в 1954 г. в возрасте 53 лет Ферми скончался от рака желудка. А в 1961 г. Фано решил вернуться к своей старой работе и опубликовал ее расширенную версию. На этот раз на английском языке в журнале \emph{Physical Review\/}~\cite{Fano:PR:1961}. В этой статье Фано применил формализм дельта-функций, что сделало работу значительно изящней.

Сегодня это одна из самых цитируемых работ из всех, когда-либо опубликованных во всех сериях \emph{Physical Review}. Чем же вызван столь ошеломляющий успех работы~\cite{Fano:PR:1961} при полном забвении работы~\cite{Fano:NC:1935}? Сам Фано по этому поводу пишет следующее~\cite{fano1977citation}: ``The paper appears to owe its success to accidental circumstances, such as the timing of its publication and some successful features of its formulation.'' --- похоже, что эта статья обязана своим успехом случайным обстоятельствам: она была опубликована в нужное время, а обсуждавшаяся в ней проблема была удачно сформулирована.

С нашей точки зрения, первая половина этой фразы не соответствует действительности, зато во второй содержатся ключевые слова --- ничего случайного в успехе статьи нет. Мало получить выдающийся результат. Нужно, чтобы этот результат был понят и воспринят коллегами. Для этого он должен быть \emph{удачно сформулирован}, а также обнародован именно в то время, когда в нем есть потребность. И сделать это нужно в правильном месте и на языке, понятном международному сообществу. Более подробное освещение этой истории можно найти в публикациях~\cite{lide2001century,berry2009_Ugo_Fano}.

\subsection{Суть резонансов Фано и их описание}

Однако нам пора вернуться от истории вопроса к обсуждению его сути. В чем же она заключается, и почему работа Фано оказалась невостребованной в 1935~г. и супервостребованной в наше время?

Начнем с опытов Бойтлера. Будем для определенности говорить о фотоионизации атома гелия. Соответствующая качественная картинка приведена на Рис.~\ref{fig:He}. Если энергия фотона, рассеиваемого атомом гелия, больше потенциала ионизации последнего, то такой фотон может быть поглощен одним из валентных электронов. В результате этот электрон приобретает энергию, превосходящую энергию связи, и покидает атом. Такой процесс фотоионизации хорошо известен каждому школьнику. Его вероятность довольно слабо зависит от частоты фотона при условии, что эта частота больше красного предела. Будем называть такое рассеяние \emph{фоновым\/} (в англоязычной литературе вместо фонового рассеяния употребляется другой термин --- background partition).

\begin{figure}
  \centering
  \includegraphics[width=\columnwidth]{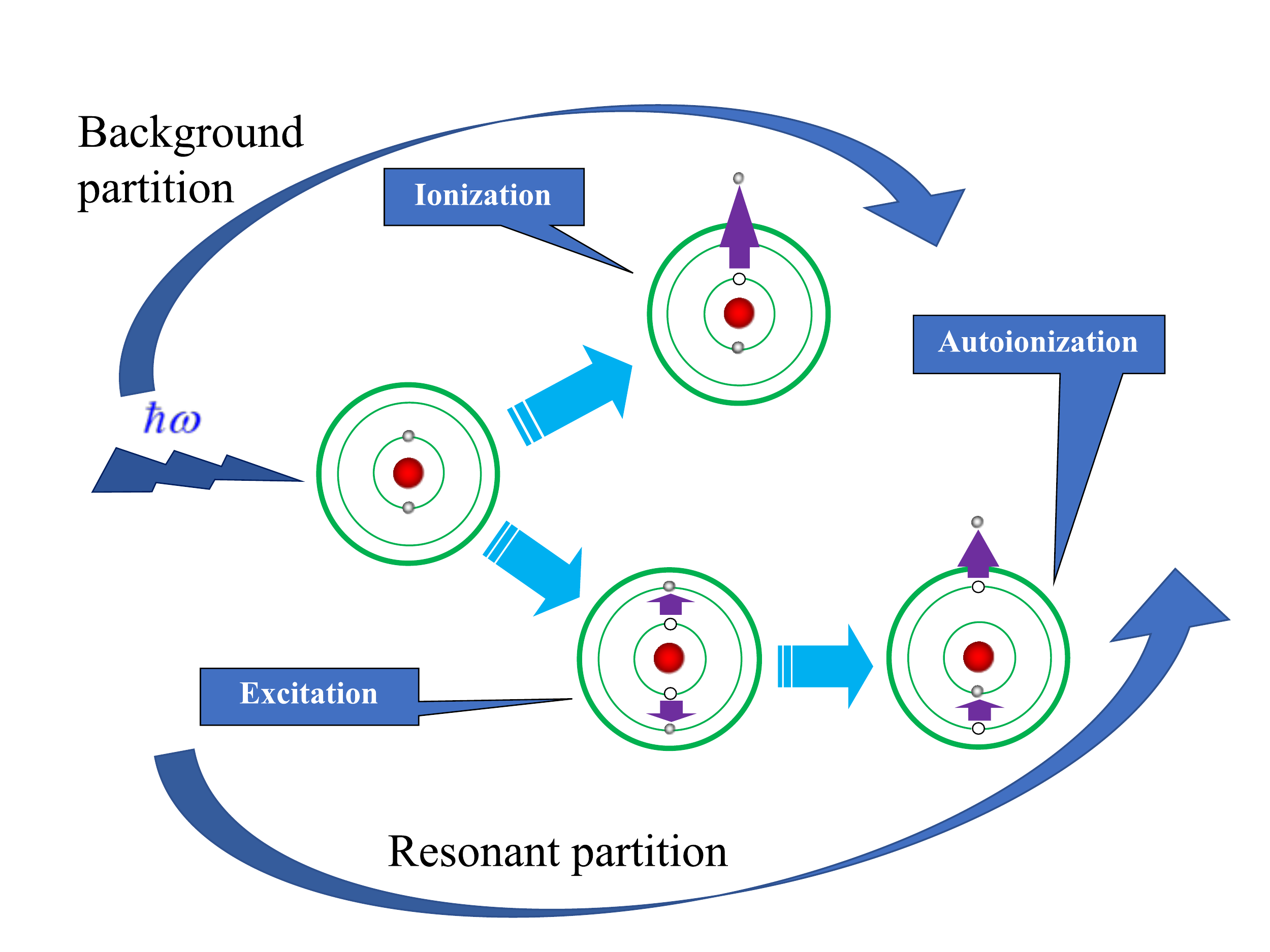}
  \caption{Две возможности для фотоионизации атома гелия. Подробности в тексте.}\label{fig:He}
\end{figure}

Однако есть и другая возможность. Если энергия фотона в точности равна сумме энергий возбуждения \emph{обоих\/} валентных электронов, но при этом меньше удвоенного потенциала ионизации, то поглощение фотона может приводить не к выбиванию из атома одного электрона, а к возбуждению сразу двух электронов. При этом оба они остаются в связанном состоянии. Дальше начинается обмен энергией между двумя электронами. Поскольку возбужденные электроны очень слабо взаимодействуют друг с другом, обмен энергии между ними идет медленно (разумеется, с точки зрения атомных временных масштабов), и атом долго живет в таком возбужденном состоянии. Но вечно в нем он пребывать не может. В конце концов один из электронов отдает свою энергию другому и переходит в основное состояние. Другой же электрон приобретает энергию, превышающую потенциал ионизации, и покидает атом. При этом ясно, что вероятность такого процесса (его называют\emph{ автоионизацией}) должна резко убывать при отклонении энергии фотона от значения, в точности равного суммарной энергии возбужденных электронов. Процесс, при котором амплитуда выходного сигнала при определенной частоте внешнего воздействия имеет резко выраженный максимум есть процесс резонансный. Поэтому рассеяние, приводящее к автоионизации, естественно также назвать \emph{резонансным\/} (resonant partition).

Как известно, в квантовой механике, если из данного начального состояния (фотон рассеивающаяся нейтральным атомом) в одно и то же конечное состояние (ионизованный атом и свободный электрон) можно прийти двумя различными путями, то каждый из таких путей характеризуется своей волновой функцией, а полная волновая функция системы есть линейная суперпозиция этих волновых функций. Суперпозиция приводит к интерференции. Интерференция же двух волн, в зависимости от разности их фаз, может давать как взаимное усиление (конструктивная интерференция), так и подавление, вплоть до полной компенсации (деструктивная интерференция).

На этом и основывалось объяснение, данное Фано экспериментам Бойтлера. Конкретные вычисления привели его к простой формуле, описывающей форму линии в этом случае:
\begin{equation}\label{eq:Fano_lineshape}
    \sigma=const\frac{( \epsilon + q)^2}{\epsilon^2+1}.
\end{equation}
Здесь $\epsilon$ --- определнным образом нормированная энергия фотона, а $q = const$ --- так называемый параметр Фано, который часто называют параметром асимметрии формы линии.

Казалось бы выражение \eqref{eq:Fano_lineshape} лишь незначительно отличается от лоренциана за счет появления в числителе $\epsilon$. В действительности же это отличие качественное. Помимо того, что оно нарушает зеркальную симметрию формы линии относительно точки $\epsilon = 0$, теперь $\sigma(\epsilon)$ обращается в ноль при $\epsilon = -q$, что, очевидно, является следствием деструктивной интерференции. Линии $\sigma(\epsilon)$ при различных значениях $q$ приведены на Рис.~\ref{fig:Fano_profiles}. При $q \rightarrow \infty$ профиль Фано сводится к обычному лоренциану.
\begin{figure}
  \centering
  \includegraphics[width=\columnwidth]{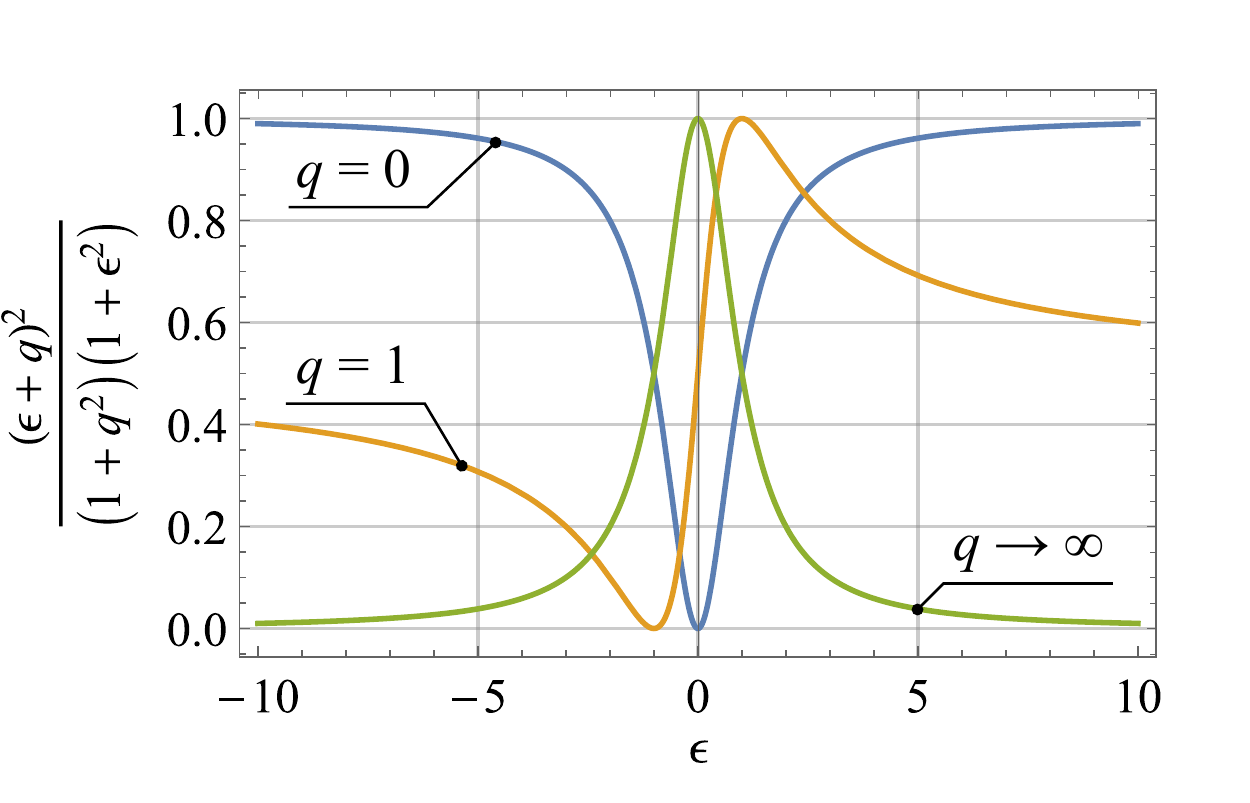}
  \caption{Профили Фано при различных значениях $q$ (указаны на рисунке). Для приведения всех профилей к одному масштабу нормировочная константа в формуле \eqref{eq:Fano_lineshape} выбрана равной $1/(1+q^2)$. Обратите внимание на резкое и почти линейное изменение профиля между точками его минимума и максимума при $q=1$.  }\label{fig:Fano_profiles}
\end{figure}

Подведем некоторые итоги. Обобщая изложенное выше, можно сказать, что резонансы Фано обусловлены неоднозначной реакцией системы на внешние воздействие (входной сигнал), порождающее не один, а два выходных сигнала, интерферирующих друг с другом. При такой формулировке становится ясным, что резонансы Фано --- это достаточно общее свойство динамических систем, которое может наблюдаться в широком классе как квантовых, так и классических задач. Более того, даже волновая природа интерферирующих сигналов не является необходимой. Вместо интерференции волн, можно рассмотреть сложение колебаний в дискретных системах. Большое количество конкретных примеров и их детальное обсуждение можно найти, например, в обзорах~\cite{Lukyanchuk:2010bt,Miroshnichenko:2010ewa} и цитированной в них литературе.

Но не только это объясняет головокружительный успех публикации Фано 1961 г. За время, прошедшее с 1935 г. до 1961 г. (и тем более --- с 1961 г. до наших дней) необыкновенно возросла точность измерений. Появилась необходимость в чувствительных сенсорах. И здесь резонансы Фано оказались необычайно полезны. Действительно, из Рис.~\ref{fig:Fano_profiles} видно, что при $q$ порядка единицы в области значений $\epsilon$, лежащих между точками минимума и максимума фановского профиля, этот профиль почти линеен и, к тому же, имеет крутой наклон. Это означает, что в данной области малое изменение входного сигнала ($\epsilon$) приводит к значительному (и, практически, линейному!) изменению выходного сигнала. Сенсор с такими свойствами --- хрустальная мечта любого специалиста, занимающегося прецизионными измерениями.

Но и это не все. Если под $\sigma$ в формуле \eqref{eq:Fano_lineshape} понимать сечение рассеяние света, то в точке $\epsilon = -q$ оно обращается в ноль. Это значит, что такая частица ничего не рассеивает, т.е. \emph{не искажает поля падающей волны}. А раз так, то она становится \emph{невидимой\/} для внешнего наблюдателя. Существенно, что при этом поле внутри частицы отнюдь не ноль, см. ниже.

Складывается парадоксальная на первый взгляд ситуация --- внутри материальной частицы возбуждено переменное электромагнитное поле, которое вызывает осцилляции свободных и/или связанных зарядов, а электромагнитного излучения, обусловленного этими осцилляциями не возникает! Такие возбуждения (их называют анаполи --- от греч. an — отрицат. частица и polos — полюс, т.е. бесполюсник) были введены в научный обиход Я. Б. Зельдовичем для объяснения нарушения четности в атомном ядре при слабых взаимодействиях~\cite{Zeld1957}. В задачах о рассеянии света нанообъектами анаполи были рассмотрены в работе~\cite{Miroshnichenko:2015gv} и в большом числе более поздних публикаций~\cite{Ospanova:LPR:2018,Yang:N:2019,Zurita-Sanchez:PRR:2019,Savinov:CP:2019,Savinov:CP:2019,Baryshnikova:AOM:2019,svyakhovskiy2019anapole,tribelsky2019dynamics,Gupta:RP:2020,Li:N:2020}.

Существование таких не излучающих возбуждений открывает широкие возможности для создания сверхточных сенсоров, способных измерять электромагнитные поля, не искажая их своим присутствием; нового поколения систем записи, считывания и обработки информации; новых материалов и пр. Здесь, однако, нас интересуют не применения этого эффекта, а вопрос, какое отношение резонансы Фано имеют к теме настоящего обзора. Попробуем в этом разобраться.

\subsection{Резонансы Фано и продольные электромагнитные моды}

Итак, если говорить о резонансах Фано при рассеянии света субволновой металлической частицей, то их отличительной особенностью должно является обращение в ноль парциального сечения рассеяния, обусловленное деструктивной интерференцией двух мод с одним и тем же значением $n$. Одинаковое значение $n$ требуется из-за того, что моды с разным значением $n$ имеют разную угловую зависимость рассеянного излучения. Поэтому их интерференция в принципе не может привести к взаимной компенсации сразу по всем направлениям. Мало того, эти интерферирующие моды должны иметь одну и ту же природу --- обе быть или электрическими или магнитными, т.к. моды разной природы даже при одном и том же значении $n$ также имеют разную угловую зависимость рассеянного излучения.

Казалось бы, все предыдущее рассмотрение говорит о том, что для малых металлических частиц таких пар мод не существует. Действительно, как следует из \eqref{eq:an_FG}, \eqref{eq:F_an_x<<1}  $a_n$ никогда не обращается в ноль, если только $\varepsilon \neq 1$. Случай же $\varepsilon = 1$ не представляет для нас интереса, т.к. при выполнении этого условия частица оптически становится тождественна окружающей среде и перестает существовать как рассеивающее волну препятствие. Из всего этого, казалось бы, можно заключить, что резонансы Фано для малых металлических частиц невозможны.

К этому моменту читатель, уже знакомый со стилем нашего обзора, должен быть уверен, что если авторы так настойчиво пытаются его в чем-то убедить, то здесь должен таиться подвох. Так и есть. Вывод, сделанный в предыдущем абзаце неверен. Резонансы Фано для малых металлических частиц возможны. Но чтобы их описать нам опять нужно выйти за пределы рассматриваемой постановки задачи.

Дело в том, что помимо $\varepsilon = 1$ в электродинамике существует еще одно выделенное значение $\varepsilon$ --- это  $\varepsilon = 0$. При таком значении диэлектрической проницаемости вектор электрической индукции $\mathbf{D}$ тождественно обращается в ноль, и становится возможным возбуждение продольных электромагнитных мод, для правильного описания которых необходимо учитывать эффекты пространственной дисперсии~\cite{LL_8}. Дисперсионная зависимость для продольных мод имеет вид:
\begin{equation}\label{eq:eps||}
  \varepsilon_{_\|}(k_{_\|},\omega)=0,
\end{equation}
в то время как поперечные моды описываются обычным дисперсионным соотношением
\begin{equation}\label{eq:eps_perp}
  k_{_{\!\perp}}^2=\varepsilon_{_{\!\perp}}\!(\omega)\omega^2/c^2.
\end{equation}
Отметим, что продольными колебания поля являются только внутри частицы --- там, где диэлектрическая проницаемость вещества близка к нулю.

Казалось бы все сложилось: колебания свободных зарядов внутри частицы, вызванные продольными  модами, должны приводить к излучению частицей электромагнитных волн, интерференция которых с излучением, обусловленным возбуждением в объеме частицы обычных поперечных колебаний поля, и создает условия для реализации резонансов Фано.

Увы, сложилось, но не все. Дело в том, что условие $\mathbf{D}=0$ сводит одно из уравнений Максвелла к условию $\nabla \times \mathbf{H}=0$. Другое же уравнение Максвелла гласит, что $\nabla \cdot \mathbf{H}=0$. Векторное поле, удовлетворяющее одновременно двум таким уравнениям, есть константа, которая в силу граничных условий на поверхности сферы должна быть равна нулю.

{Таким образом, рассматриваемые нами продольные колебания поля являются не электромагнитными, а электрическими --- магнитное поле при таких колебаниях не возбуждается. Раз так, то вектор Пойнтинга таких колебаний тождественно равен нулю, и следовательно, в силу непрерывности потока энергии через границу сферы, продольные моды являются \emph{неизлучающими\/} и никакого вклада в интерференцию внести не могут. }

{Тут обиженный читатель вправе высказать законный упрек авторам, отвлекающих его внимание на обсуждения вопросов, может быть и занятных, но не имеющих отношения к рассматриваемой проблеме. Не спеши, дорогой читатель, упрекать авторов. Имеют, имеют продольные моды отношение к рассматриваемой проблеме. Сами они действительно ничего не излучают, но, связанные с ними колебания электрического поля через граничные условия зацепляются с обычными поперечными модами. А вот эти-то поперечные моды прекрасно излучают. Они-то и вносят в интерференционную картину необходимый для реализации резонансов Фано вклад.  }

Здесь также следует подчеркнуть два важных обстоятельства. Во-первых, обычно, в отличие от временной дисперсии (зависимости $\varepsilon$ от $\omega$), пространственная дисперсия --- слабый эффект, которым, как правило, можно пренебречь~\cite{LL_8}. В данном случае он выходит на первую роль только потому, что в главном приближении (без учета пространственной дисперсии) диэлектрическая проницаемость обнуляется. Как всегда в таких случаях, когда главный член обращается в ноль, необходимо учитывать поправки к нему, которые в обычных обстоятельствах пренебрежимо малы.

Второе обстоятельство связано с тем, что учет пространственной дисперсии означает, что зависимость вектора электрической индукции $\mathbf{D}$ от вектора электрического поля $\mathbf{E}$ становится нелокальной --- помимо значения  $\mathbf{E}$ в данной точке он начинает зависть и от значения производных от $\mathbf{E}$ по пространственным координатам~\cite{LL_8}. Это повышает порядок уравнений Максвелла, так что для правильной постановки соответствующей краевой задачи требуются дополнительные граничные условия.

\begin{figure}
  \centering
  \includegraphics[width=.87\columnwidth]{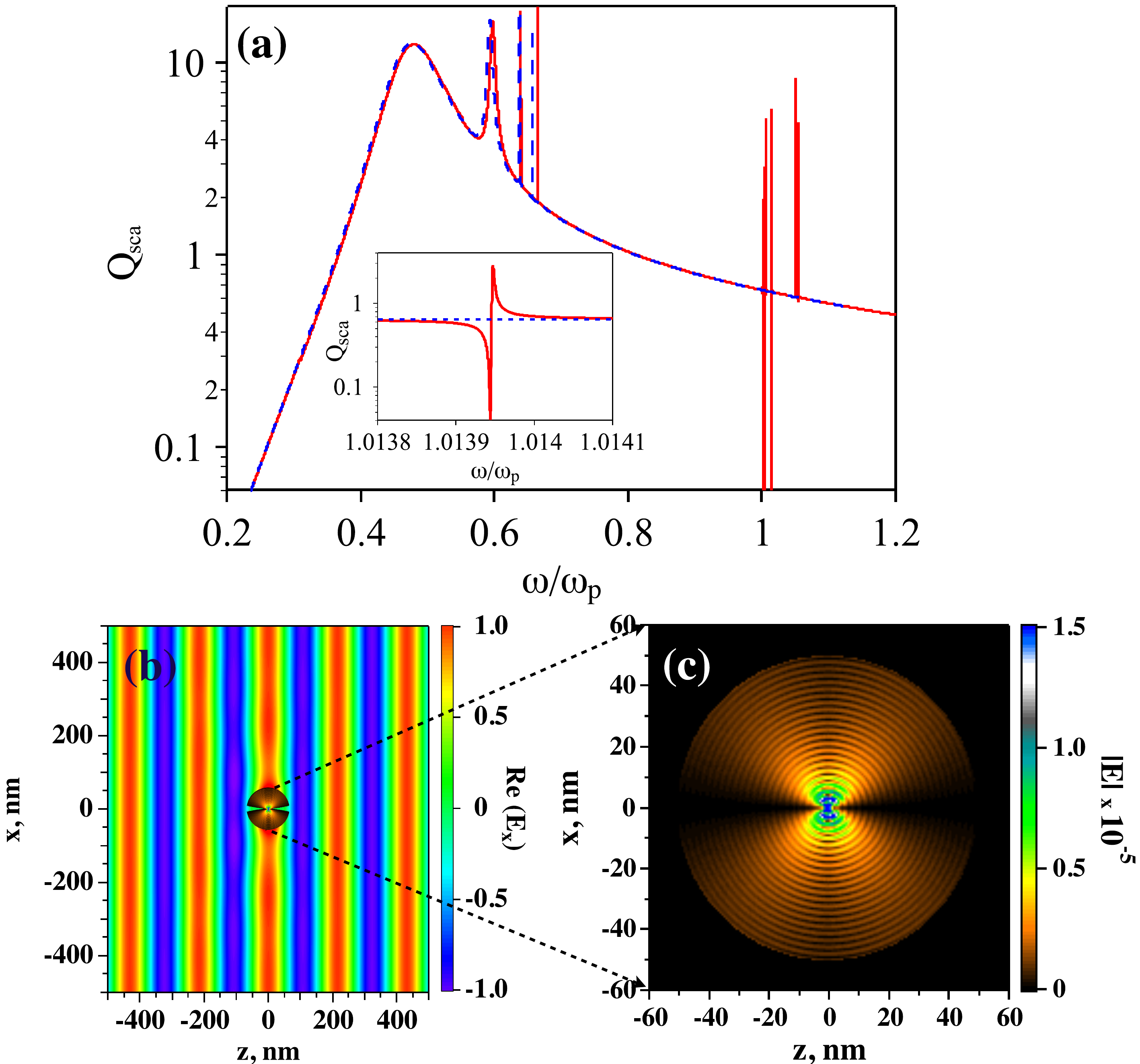}
  \caption{Рассеяние плоской линейно поляризованной электромагнитной волны металлической сферой (натрий) радиусом 50 нм \cite{Tribelsky:2012ju}. (а) --- Зависимость эффективности рассеяния от частоты. Сплошная линия - с учетом вклада от продольных электромагнитных мод, пунктир --- традиционная теория Ми (без учета такого вклада). Врезка показывает форму одной из линий резонансов Фано. (b) --- Профиль пространственной зависимости электрического поля в произвольный фиксированный момент времени. Частота падающего излучения соответствует точке деструктивного резонанса на врезке в панель (а). Вектор $\mathbf{E}$ нормирован на амплитуду электрического поля в падающей волне и лежит в плоскости $xz$. Падающая волна распространяется вдоль положительного направления оси $z$. Частица практически не искажает поля падающей волны, т.е. почти совершенно невидима. (с) --- Профиль электрического поля внутри частицы. Обратите внимание, что масштаб цветовой шкалы на панеле (с) в $10^5$ раз больше чем на панели (b). Этот же масштаб использован на панели (b) для изображения поля внутри частицы.}\label{fig:Ruppin}
\end{figure}

Вопрос о формулировке этих дополнительных условий оказывается непростым и до сих пор до конца нерешенным, см.~\cite{Golubkov:ABC:UFN}. Решение обобщенной задачи Ми, учитывающей возможность возбуждения в объеме частицы продольных электромагнитных мод при различном выборе дополнительных граничных условий, обсуждалось в работе~\cite{ruppin1981_diffeterntABC}. В ней показано, что хотя положение резонансных частот зависит от конкретного выбора граничных условий, качественно структура спектра оказывается к нему инвариантна. Поэтому для понимания существа дела достаточно ограничиться рассмотрением одного конкретного случая такого выбора.

Соответствующее исследование было проведено в работе~\cite{Tribelsky:2012ju}, где в качестве дополнительного граничного условия использовалось условие непрерывности на границе сферы нормальной компоненты тока смещения $\frac{1}{4\pi}\frac{\partial \mathbf{E}}{\partial t}$~\cite{ruppin1975}, в качестве зависимости $\varepsilon_{_{\!\perp}}(\omega)$ --- формула Друде \eqref{eq:epsilon_Drude}, а в качестве дисперсионной зависимости $\varepsilon_{_{\|}}(k_{_{\|}}\omega)$ --- феноменологическое соотношение
\begin{equation}\label{eq:e||_phenom}
 \varepsilon_{_{\|}}(k_{_{\|}}\omega) = \varepsilon_{_{\!\perp}}\!(\omega)-\alpha k^2_{_{\|}}R^2,
\end{equation}
справедливое в пренебрежении анизотропией материала, из которого изготовлена рассеивающая излучение частица. Здесь $\alpha$ --- безразмерная феноменологический коэффициент по порядку величины равный отношению $d^2/R^2$, где $d$ --- межатомное расстояние. Конкретно в \cite{Tribelsky:2012ju} использовалось следующие выражение для $\alpha$:
\begin{equation}\label{eq:alpha}
{\alpha = \frac{3}{5}\left(\frac{{v}_{_F} \omega_p}{\omega^2 R}\right)^2,}
\end{equation}
где ${v}_{_F}$ --- фермиевская скорость свободных электронов. При этом были выбраны такие значения параметров: $\omega_p= 8.65\times10^{15}$ с$^{-1}$, ${v}_{_F} = 1.07 \times 10^8$~см/с, что примерно соответствует натрию, и $R=50$~нм.

Проделанный в работе~\cite{Tribelsky:2012ju} анализ, показал, что учет вклада продольных электромагнитных колебаний приводит к образованию в окрестности $\omega = \omega_p$ каскада резонансов Фано, имеющих чрезвычайно узкую ширину линии и характеризующихся гигантской концентраций поля внутри рассеивающей частицы, в том числе и в точках деструктивной интерференции, см. Рис.~\ref{fig:Ruppin}. К сожалению, как это всегда бывает при резонансах высокой добротности, эффект оказывается весьма чувствительным к диссипативному затуханию и быстро подавляется с ростом последнего.
\subsection{Направленные резонансы Фано}
\subsubsection{Теория}

Вернемся к обсуждению природы резонансов Фано. Для их реализации при рассеянии волн необходимо иметь по крайней мере две интерферирующих рассеянных волны. При этом, для возникновения деструктивной интерференции эти две волны должны одновременно удовлетворять двум условиям: иметь равные амплитуды и противоположные фазы.

Обратимся опять к выражению для коэффициентов рассеяния \eqref{eq:an_FG}. Как уже отмечалось, в недиссипативном пределе функции $F_n$ и $G_n$ чисто действительны. При этом на крыльях резонансной линии, когда $|G_n| \gg |F_n|$, $a_n \approx -iF_n/G_n$ и является почти чисто мнимой величиной.

Далее, по определению, в точке резонанса функция $G$ обращается в ноль. Следовательно при прохождении точки резонанса она меняет знак, см. \eqref{eq:G_an_x<<1}. Иными словами, при прохождении резонанса фаза коэффициента рассеяния меняется на $\pi$ --- факт хорошо известный из теории колебаний.

Имея все это ввиду, рассмотрим Рис.~\ref{fig:Q_Drude}. На этом рисунке изображена сравнительно широкая линия дипольного резонанса, крыло которой буквально прошивает интенсивная и очень узкая линия квадрупольного резонанса. При этом в точке квадрупольного резонанса должно выполняться условие $\varepsilon \approx - 3/2$. Тогда, в этой области  парциальная эффективность дипольного рассеяния порядка $x^{2} \ll 1$, а парциальная эффективность квадрупольного резонанса в максимуме порядка $1/x^2 \gg 1$, см. \eqref{eq:sigma_sca_n_max}--\eqref{eq:G_an_x<<1}.

Из этого следует, что эффективности дипольного и квадрупольного рассеяния (а следовательно и соответствующие поля парциальных волн) оказываются равны в двух точках, лежащих по обе стороны от точки максимума квадрупольного рассеяния. При этом в точках равенства амплитуда каждой из парциальных мод все еще сильно меньше максимального значения амплитуды квадрупольной моды в точке резонанса. Иными словами, обе точки равенства лежат на крыльях линии квадрупольного резонанса. Сама же область квадрупольного резонанса лежит на крыле дипольной линии. Тогда с учетом сказанного в предыдущем параграфе заключаем, что в одной из точек равенства амплитуд фазы коэффициентов рассеяния $a_1$ и $a_2$ равны, а в другой противоположны. Иначе говоря, \emph{в окрестности одной из этих точек с неизбежностью должны выполняться оба условия деструктивной интерференции}.

Однако такая деструктивная интерференция качественно отличается от обсуждавшегося ранее случая. В традиционных резонансах Фано интерферируют моды, имеющие одну и ту же угловую зависимость рассеянного излучения. Поэтому равенство амплитуд и противоположность фаз этих мод обеспечивает обращение в ноль интенсивности рассеянного излучения сразу во всех направлениях, т.е. обнуление соответствующего \emph{интегрального\/} сечения рассеяния. Сейчас же речь идет об интерференции мод с \emph{разными\/} значениями $n$. Как уже отмечалось, для таких мод рассеянное поле имеет разную угловую зависимость. Поэтому обращение в ноль интенсивности рассеянного излучения сразу во всех направлениях за счет их деструктивной интерференции принципиально невозможно. Речь может идти только об обнулении интенсивности рассеяния вдоль определенного направления, т.е. \emph{дифференциального\/} сечения рассеяния, для которого при данной частоте в точности выполняется равенство амплитуд и противоположность фаз интерферирующих волн. При этом для разных направлений эти частоты будут различны.

Таким образом, интерференция полей, создаваемых разными мультиполями, приводит к тому, что интенсивность излучения, рассеянного вдоль данного фиксированного направления, будучи измеренная как функция частоты падающего излучения, имеет асимметричный профиль Фано. В то же время, в соответствии с \eqref{eq:sigma_sph_n}, \eqref{eq:an&gamma_n_x<<1}, интегральное сечение рассеяния будет описываться симметричным лоренцианом~\cite{Tribelsky:2008dha}.

Для того, чтобы количественно описать эти эффекты нам осталось рассмотреть интерференцию полей, создаваемых различными мультиполями. Это легко сделать, воспользовавшись известным представлением интенсивности излучения, рассеянного вдоль данного направления. Если выбрать сферическую систему координат, центр которой совпадает с центром рассеивающей сферы, то для плоской линейно поляризованной падающей волны, распространяющейся вдоль положительного направления оси $z$, с плоскостью колебаний вектора $\mathbf{E}$, совпадающей с плоскостью $xz$, интенсивность, рассеянная вдоль данного направления имеет в дальней волновой зоне следующие компоненты вектора $\mathbf{E}$~\cite{Born_Wolf:Optics}:
\begin{equation}\label{eq:E_sca}
  |E^{\rm sca}_\theta|^2 = I_{_\|}\!\cos^2\varphi,\;\; |E^{\rm sca}_\varphi|^2 = I{_{\!_ \perp}}\!\sin^2\varphi,
\end{equation}
где $\theta$ и $\varphi$ --- полярный и азимутальный углы, соответственно, а $I_{_\|}$ и $I{_{\!_ \perp}}$ определяются соответствующим мультипольным разложением. Согласно сказанному, в окрестности квадрупольного резонанса ($\varepsilon = -3/2$) из этого разложения, нам достаточно оставить только дипольный и квадрупольный члены. При этом, в главном приближении в дипольном члене мы можем положить $\varepsilon = -3/2$, а в квадрупольном --- воспользоваться формулой \eqref{eq:an&gamma_n_x<<1}.

В результате, с точностью до несущественного сейчас общего для $I_{_{\|,\perp}}$ множителя, получаем~\cite{Tribelsky:2008dha}:
\begin{eqnarray}
  I^{(s)}_\parallel &\propto& \left| i x^3 \cos\theta +
    \frac{x^5\cos2\theta}{2(x^5 -12i\delta\varepsilon)}\right|^2,\label{eq:Ipar} \\
  I^{(s)}_\perp &\propto& \left| i x^3 +
    \frac{x^5\cos\theta}{2(x^5 -12i\delta\varepsilon)}\right|^2.\label{eq:Iper}
\end{eqnarray}

Отметим, что, в полном соответствии с обсуждавшейся выше качественной картиной, полученные выражения имеют два характерных масштаба. Один из них описывает узкий резонансный пик, связанный с квадрупольным резонансом. Его ширина определяется формулой \eqref{eq:an&gamma_n_x<<1} для $\gamma_n$ при $n=2$ и равена $x^5/6$. Другой, значительно более широкий, определяет точку деструктивного резонанса и несколько различен для двух поляризаций рассеянного света:  $\delta\varepsilon_{_\|} = -x^2\cos2\theta/24\cos\theta$ и $\delta\varepsilon_{\!_\perp} = -x^2\cos\theta/24$.

Примечательно, что согласно \eqref{eq:Ipar}, \eqref{eq:Iper} интенсивности $I_{_{\|, \perp}}$ зависят от $\theta$, но не зависят от $\varphi$, т.е. остаются постоянными на поверхности конусов с осью, совпадающей с осью $z$ и фиксированным углом раствора, но меняются при изменении угла раствора. Подчеркнем также, что в зависимости от величины $\theta$ и поляризации рассеянного излучения точки деструктивной интерференции могут находиться по разные стороны от точки $\delta\varepsilon = 0$, определяющей положение максимума квадрупольного резонанса. Таким образом, направленные резонансы Фано являются мощным инструментом, позволяющим создавать различные диаграммы направленности рассеянного излучения, существенно подавляя рассеяния вдоль практически любого желаемого направления только за счет надлежащего выбора частоты падающей волны.

Это в теории. Что в данном случае говорит <<его величество эксперимент>>? Надо сказать, что несмотря на идейную простоту, технически проведение экспериментов по кругу вопросов, обсуждаемых в этом обзоре, как правило представляет весьма сложную задачу, связанную с необходимостью проведения прецизионных измерений на наномасшабах. По-видимому, по этой причине, несмотря на то, что теоретически направленные резонансы Фано обсуждались еще в 2008~г.~\cite{Tribelsky:2008dha,OPN:2008}, их экспериментальное подтверждение было получено только восемь лет спустя~\cite{PRB:DirectionaFano2016}.

\subsubsection{Эксперимент}
\begin{figure}
  \centering
  \includegraphics[width=\columnwidth]{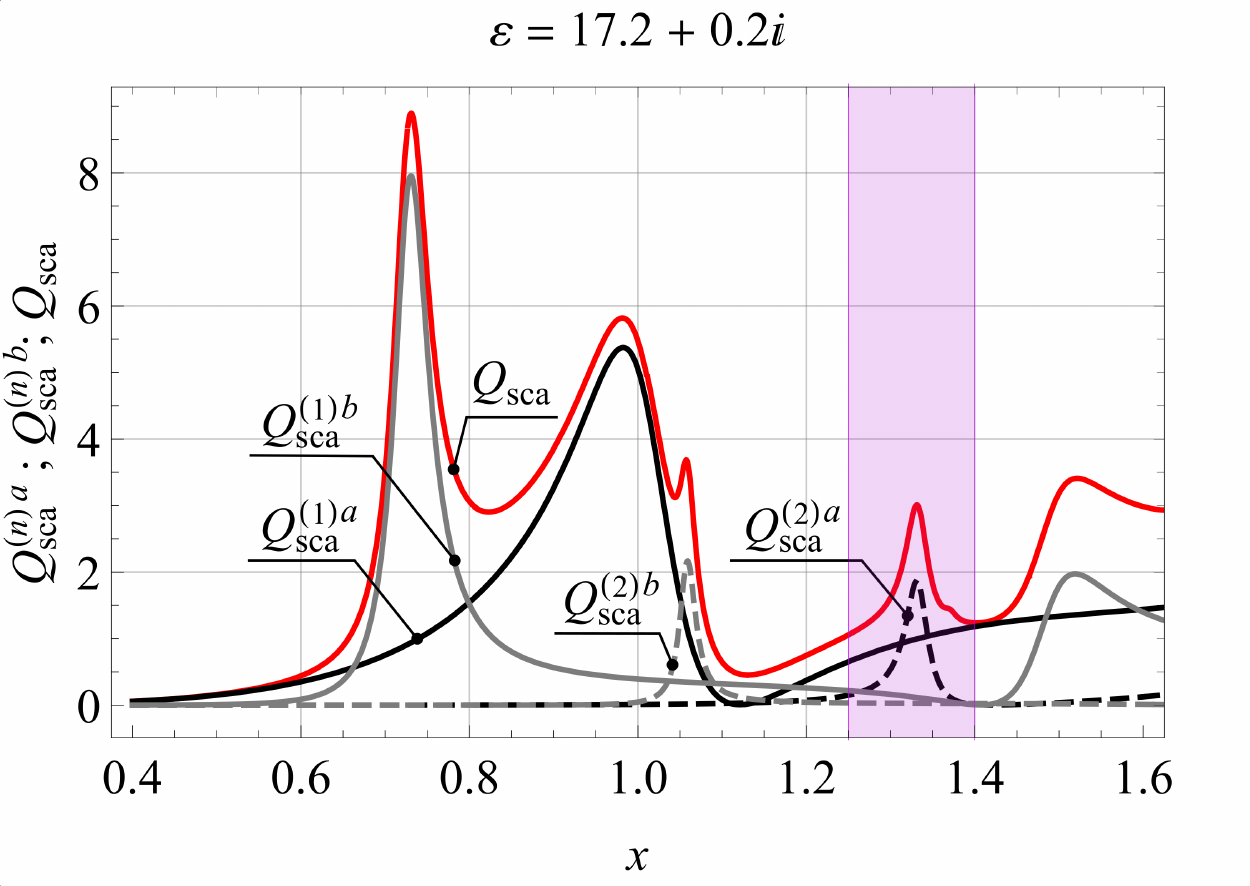}
  \caption{Интегральная эффективность рассеяния $Q_{\rm sca}$ и парциальные эффективности четырех первых мультиполей: электрического диполя ($Q^{(1)\,a}_{\rm sca}$), магнитного диполя ($Q^{(1)\,и}_{\rm sca}$), электрического квадруполя ($Q^{(2)\,a}_{\rm sca}$), и магнитного квадруполя ($Q^{(2)\,b}_{\rm sca}$), для сферической частицы с $\varepsilon=17.2+0.2i$, как функции ее параметра размера. Область, где можно наблюдать направленные резонансы Фано, отмечена заливкой. }\label{fig:Q_17}
\end{figure}

Перейдем к краткому обсуждению этих экспериментов. Прежде всего, отметим, что направленные резонансы Фано реализуются при перекрытии крыла относительно широкой линии мультиполя с одним значением $n$ с узкой резонансной линией моды другой мультиполярности. Далее, координатная зависимость излучения каждого мультиполя определяется видом собственных функций задачи. Она зависит от симметрии задачи, но не зависит от материальных свойств рассеивающей частицы. Последние влияют только и исключительно на значения коэффициентов рассеяния.

\begin{figure}
  \centering
  \includegraphics[width=.95\columnwidth]{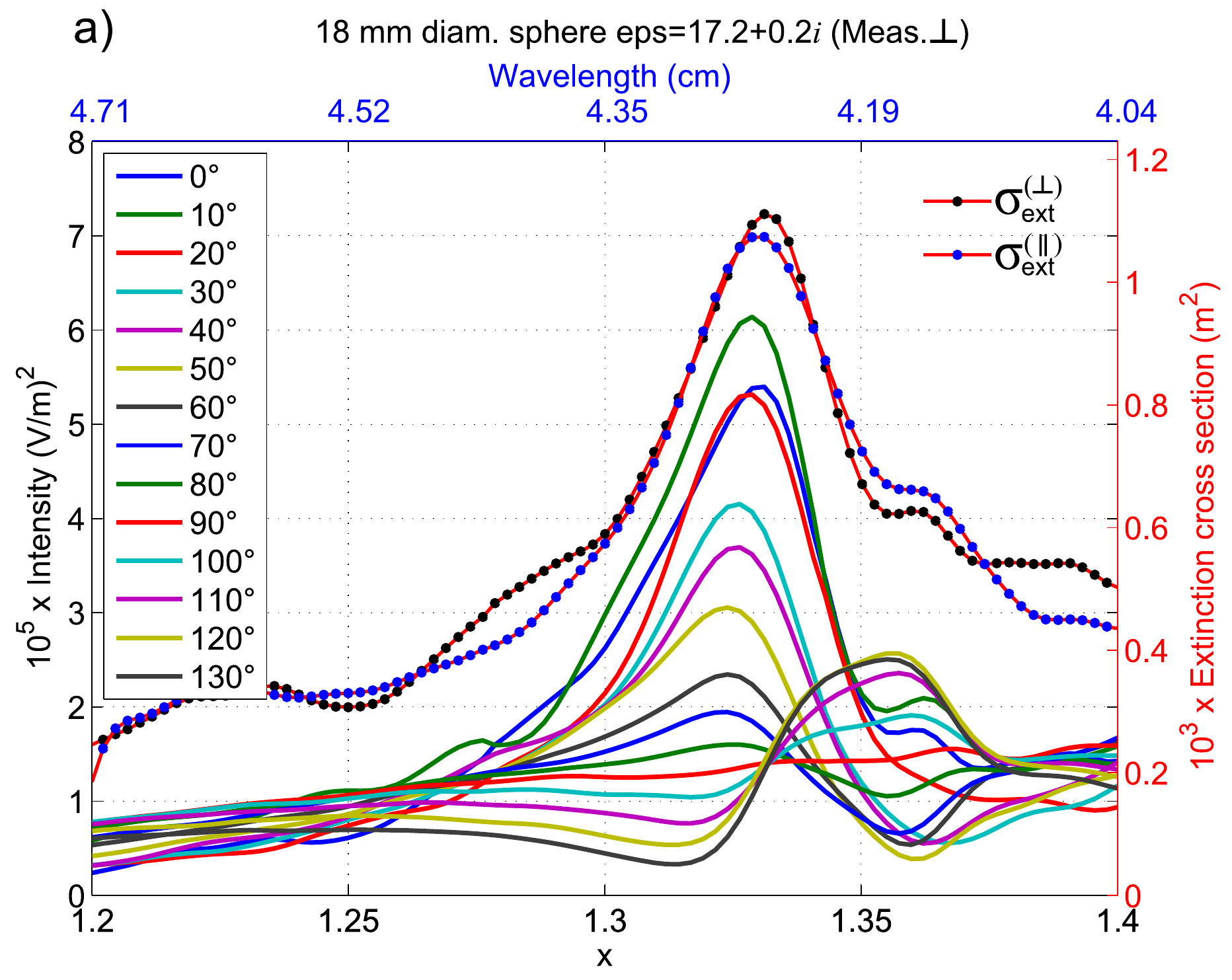}\\
  \includegraphics[width=.95\columnwidth]{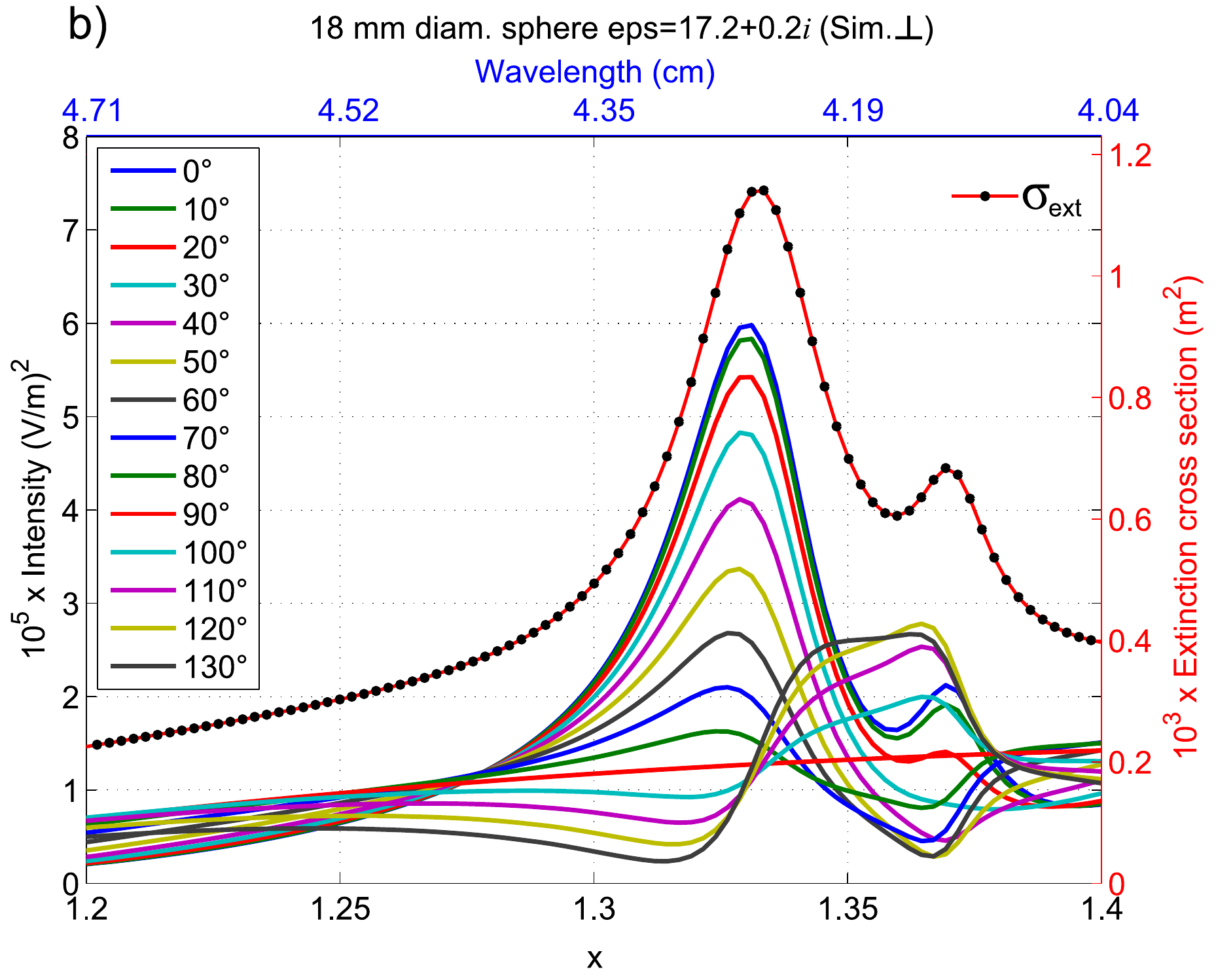}
  \caption{Интенсивность рассеяния $I_{\!{_\perp}}$ вдоль различных направлений, определяемых значениями полярного угла $\theta$ (указаны на врезке), как функция длины волны излучения. Керамическая сфера диаметром 18 мм. В указанном диапазоне дисперсионной зависимостью $\varepsilon(\omega)$ можно пренебречь и считать $\varepsilon(\omega) \approx 17,2 + 0,2i$. (a) --- результаты измерений, (b) --- расчет по формулам Ми. Детали в тексте.}\label{fig:Geffrin}
\end{figure}

Из сказанного ясно, что отрицательность действительной части диэлектрической проницаемости рассеивающей частицы вовсе не является обязательным свойством, необходимым для наблюдения направленных резонансов. Диэлектрики, для которых Re$\,\varepsilon >0$, также могут демонстрировать такие резонансы при условии, что спектральная зависимость их резонансных линий удовлетворяет отмеченным выше свойствам.

Другим важным обстоятельством является масштабная инвариантность уравнений Максвелла. Если изменить геометрические размеры задачи, одновременно с длиной волны излучения так, что их отношение останется неизменным, а диэлектрическую проницаемость оставить неизменной, то такая задача рассеяния окажется идентична исходной. Это позволяет моделировать рассеяние света наночастицей, изучая эквивалентную задачу по рассеянию существенно более длинноволнового излучения макроскопическим объектом той же формы.

Такое моделирование и было использовано авторами работы~\cite{PRB:DirectionaFano2016} для экспериментального исследования направленных резонансов Фано. В частности, из специальной керамики, имеющей в сантиметровом диапазоне $\varepsilon \approx 17.2 + 0.2i$, что примерно соответствует диэлектрической проницаемости широко распространенных полупроводниковых материалов (Si, Ge, GaAs, GaP) в оптическом и ближнем ИК диапазонах, был изготовлен шар диаметром 18 мм. Интегральная эффективность рассеяния такого шара и парциальные эффективности его дипольных и квадрупольных мод представлена на Рис.~\ref{fig:Q_17}, как функции его параметра размера. Видно, что в области $1,25 < x < 1,40$ определяющий вклад в полную эффективность рассеяния вносят электрическая дипольная и квадрупольная моды, и что в этой области выполнены условия для наблюдения направленных резонансов Фано, связанных с интерференцией излучения этих двух парциальных мод.

Для исследования угловой и частотной зависимости рассеянного излучения керамическая сфера помещалась в специальную камеру с безотражательным покрытием ее стенок, где она облучалась плоской, линейно поляризованной волной сантиметрового диапазона. Излучение, рассеянное сферой вдоль данного направления, принималось приемной антенной. Сечение экстинкции определялось на основании оптической теоремы~\cite{Bohren::1998,Born_Wolf:Optics} по измерению амплитуды и фазы рассеянной вперед волны при двух независимых поляризациях излучения.

Мы не будем здесь останавливаться на деталях этого эксперимента, достаточно сложных в техническом отношении, отсылая заинтересованного читателя к оригинальной работе~\cite{PRB:DirectionaFano2016} и цитированной в ней литературе. В качестве примера приведем взятые из этой работы графики зависимости интенсивности $I_{\!{_\perp}}$, рассеянной вдоль данного направления, от длины волны излучения при различных значения полярного угла $\theta$, см. Рис.~\ref{fig:Geffrin}(а). На Рис.~\ref{fig:Geffrin}(b) приведены те же зависимости, рассчитанные на основании точного решения Ми. Подчеркнем, что при построении графиков не было использовано ни одного подгоночного параметра. Отметим также, при переходе значений $\theta$ через точку $\theta = 90^\circ$ взаимная последовательность точек минимума и максимума профиля $I_{\!{_\perp}}(x)$ меняется на противоположную, см.~\eqref{eq:Iper}. Обращает на себя внимание тот факт, что при $\theta > 90^\circ$ локальный минимум рассеяния в данном направлении располагается почти при том же значении $x$, которому соответствует локальный максимум интегрального сечения экстинкции. Этот результат трудно было бы понять вне рамок развитых здесь представлений.

\section{Заключение}

Подведем некоторые итоги. В данном обзоре мы сумели обсудить только несколько новых и, с нашей точки зрения, достаточно необычных аспектов старой проблемы рассеяния света малыми частицами. Сохраняя объем обзора в разумных пределах, мы пожертвовали рядом важных вопросов и приложений ради возможности более подробно обсудить физику основных процессов и явлений, с которыми связаны эти новые аспекты. К таким исключенным из обсуждения вопросам прежде всего относятся проблемы рассеяния света частицами сложной формы (диски; эллипсоиды, в том числе сильно вытянутые или сильно сплюснутые; частицы тороидальной формы --- <<бублики>>; пирамиды и другие могогранники и т.п.), кластерами из нескольких наночастиц --- <<метамолекулами>>, а также метаматериалами и метаповерхностями. Их обсуждению посвящено большое число публикаций как оригинальных (включающих и экспериментальные, и теоретические работы), так и обзоров и монографий, см., например, работы~\cite{Klimov:Nanoplasmonika,Mishchenko2000,Papasimakis:NM:2016,Rahmani2013,shalaev2013planar,Yu2014,cui2014coding,Kabashin2019,Feng2020}, к каковым мы и адресуем заинтересованного читателя.

Мы считаем оправданным такой отбор материала, т.к. глубокое понимание физики рассмотренных в обзоре относительно простых случаев помогает составить общую картину явления, характерную и для более сложных задач. Мало того, в ряде случаев обсуждаемое в обзоре решение Ми для сферы может быть количественно использовано как базисное приближение для описания таких более сложных случаев (сфера на подложке, метамолекулы и др.) различными методами теории возмущений.

Остановимся теперь кратко на возможных применениях обсуждаемых эффектов. Область таких применений весьма велика. В частности, резонансное локальное усиление электромагнитного поля позволяет создавать участки субволнового размера для избирательного мощного электромагнитного воздействия на биологические ткани, что открывает принципиально новые возможности в генной инженерии и нанохирургии на клеточном уровне, для лечения онкологических заболеваний, создания новых лекарственных препаратов и пр.

Эти же свойства, дополненные возможностями свербыстрого переключения между различными состояниями и управлением диаграммами рассеяния, могут быть широко использованы в разработки новых методов передачи, записи и хранения информации сверхвысокой плотности, а также ее сверхбыстрой обработки, включая создание принципиально новых многофункциональных элементов интегральной оптики со сверхплотной упаковкой. Последние могут быть использованы для создания сверхпроизводительных оптических компьютеров, построенных на принципах волоконной оптики. Данные свойства могут также послужить основой для прорыва в области визуализации объектов со сверхвысоким пространственным и временным разрешением, а также в спектроскопии сверхвысокого разрешения.

В обзоре мы совершенно не касались обсуждения различных нелинейный эффектов. Однако отмечавшийся выше эффект воронки указывает на возможность создания областей с громадной плотностью электромагнитной энергии при весьма умеренных значениях плотности потока падающего излучения. Если при этом рассеивающая частица помещена в нелинейную среду (или сама обладает нелинейными свойствами), то наличие таких областей должно приводить к рекордному снижению порогов генерации высших гармоник, что в свою очередь, открывает перспективы для создания нового поколения биологических и медицинских маркеров. Такие маркеры позволят получить сигнал отклика со сверхвысоким (субклеточном) пространственным разрешением при рекордно низких значениях интенсивности зондирующего импульса и соответственного снижения его действия на исследуемые биологически ткани.

И это только некоторые из примеров. Список легко продолжить. Мы надеемся что данный обзор, может быть полезен при проведении такого рода исследований.

Авторы благодарны А. С. Сигову за внимательное прочтение рукописи и полезные замечания, способствующие ее улучшению. Обзор написан при финансовой поддержке РФФИ (научный проект \mbox{№ 19-12-50016)} и проекта РНФ \mbox{№ 19-72-30012,} в рамках которого выполнены все оригинальные расчеты, приведенные в настоящей публикации.

\section{Математическое дополнение}
\subsection{Точное решение Ми для сферы}

Приведем здесь для справок основные результаты теории Ми. Рассмотрим рассеяние однородной сферой радиуса $R$, центр которой помещен в начало сферической системы координат, плоской линейно поляризованной электромагнитной волны с циклической частотой $\omega$ и волновым вектором $\mathbf{k}$ ориентированным вдоль оси $z$, так что электрическое поле такой волны описывается формулой
\begin{equation}\label{eq:Eincident}
 \mathbf{E}^{\rm inc}=\mathbf{E_0}e^{i(kz-\omega t)},
\end{equation}
где вектор $\mathbf{E}$ колеблется вдоль оси $x$. Удобно ввести безразмерные поля $\mathcal{E}=E/E_0$ и $\mathcal{H}=H/H_0$. Здесь \mbox{$H_0 = E_0\sqrt{\varepsilon_{\rm out}/\mu_{\rm out}}$} --- амплитуда магнитного поля в падающей волне. Для большей общности сейчас не предполагается, что магнитная проницаемость тождественно равна единице. Это позволяет использовать приведенные ниже формулы и для магнитных частиц. Разумеется, в этом случае относительный коэффициент преломления сферы $m$ следует считать равным \mbox{$m=\sqrt{\frac{\varepsilon_{\rm in}\mu_{\rm in}}{\varepsilon_{\rm out}\mu_{\rm out}}}$}.

Разлагая поле падающей волны по сферическим гармоникам, его можно представить в виде бесконечного ряда парциальных волн~\cite{Kerker2013,Bohren::1998,Bohren::1998,Born_Wolf:Optics}:
\begin{equation}\label{eq:EH_incident}
  \mathcal{E}^{\rm inc}_{r,\theta,\varphi} = e^{-i\omega t}\sum_{n=1}^{\infty} \mathcal{E}^{{\rm inc}\,(n)}_{r,\theta,\varphi};\; \mathcal{H}^{\rm inc}_{r,\theta,\varphi} = e^{-i\omega t}\sum_{n=1}^{\infty} \mathcal{H}^{{\rm inc}\,(n)}_{r,\theta,\varphi},
\end{equation}
где

\begin{eqnarray}
    \mathcal{E}^{{\rm inc}\,(n)}_{r} &=& i^{n-1}\frac{2n+1}{(kr)^2}\psi_n(kr)\pi_n(\theta)\sin\theta\cos\varphi, \label{eq:Einc_r}\\
    \mathcal{E}^{{\rm inc}\,(n)}_{\theta} &=& i^n\frac{2n+1}{kr(n+1)n}\bigg[\psi_n(kr)\pi_n(\theta)- \label{eq:Einc_theta}\\
    & &-i\psi'_n(kr)\tau_n(\theta)\bigg]\cos\varphi, \nonumber\\
    \mathcal{E}^{{\rm inc}\,(n)}_{\varphi} &=& -i^n\frac{2n+1}{kr(n+1)n}\bigg[\psi_n(kr)\tau_n(\theta)-\label{eq:Einc_phi}\\
    & &-i\psi'_n(kr)\pi_n(\theta)\bigg]\sin\varphi, \nonumber\\
    \mathcal{H}^{{\rm inc}\,(n)}_{r} &=&
    i^{n-1}\frac{2n+1}{(kr)^2}\psi_n(kr)\pi_n(\theta)\sin\theta\sin\varphi,
\end{eqnarray}
\begin{eqnarray}
    \mathcal{H}^{{\rm inc}\,(n)}_{\theta} &=& i^n\frac{2n+1}{kr(n+1)n}\bigg[\psi_n(kr)\pi_n(\theta)- \label{eq:Einc_theta}\\
    & &-i\psi'_n(kr)\tau_n(\theta)\bigg]\sin\varphi, \nonumber\\
    \mathcal{H}^{{\rm inc}\,(n)}_{\varphi} &=& i^n\frac{2n+1}{kr(n+1)n}\bigg[\psi_n(kr)\tau_n(\theta)-\label{eq:Einc_phi}\\
    & &-i\psi'_n(kr)\pi_n(\theta)\bigg]\cos\varphi. \nonumber
\end{eqnarray}
Здесь $\theta$  --- полярный, а $\varphi$ --- азимутальный угол,  штрих означает производную по всему аргументу функции и введены следующие обозначения:
\begin{eqnarray}
\!\!\!\!\!\! \psi_n(z) &=& \sqrt{\frac{\pi z}{2}}J_{n+\frac{1}{2}}(z),\; \pi_n(\theta)=\frac{P_n^1(\cos\theta)}{\sin\theta},\label{eq:notation_psi_pi_n}\\
\!\!\!\!\!\! \tau_n(\theta) &=& \frac{dP_n^1(\cos\theta)}{d\theta},\; P_n^m(z)\!=\!(1-z^2)^{\frac{m}{2}}\frac{d^mP_n(z)}{dz^m},\label{eq:notation_tau_P_mn}
\end{eqnarray}
где $J_{n+\frac{1}{2}}(z)$ - функция Бесселя, а $P_n(z)$ - полиномы Лежандра. Функцию $\psi_n(z)$ и введенную ниже $\xi_n(z)$ часто называют функциями Риккати-Бесселя.

При выбранной симметрии задачи --- рассеяния сферической частицей плоской электромагнитной волны \eqref{eq:Eincident} в ее разложение по сферическим гармоникам входят только моды с $m=1$ (здесь $m$ --- верхний индекс присоеденненных полиномов Лежандра в выражении \eqref{eq:notation_tau_P_mn}, не путать с коэффициентом преломления, обозначаемым той же буквой). В других случаях, например, при возбуждении сфокусированными или сингулярными пучками данная симметрия нарушается и нужно учитывать все гармоники \cite{Maheu:JO:1988,Gouesbet:JO:1988,Nieminen:JQSRT:2003}. Поскольку, однако, в силу полноты системы плоских волн произвольную электромагнитную волну можно представить как их суперпозицию, мы такие случаи обсуждать не будем, сосредоточившись на рассеянии плоской монохроматической волы.

Поле рассеянного излучения (индекс sca) и поле внутри частицы (индекс in) представляются рядами, аналогичными \eqref{eq:EH_incident}, где соответствующие парциальные волны определяются выражениями:

\begin{eqnarray}
    \mathcal{E}^{{\rm sca}\,(n)}_{r} &=& i^{n+1}\frac{2n+1}{(kr)^2}a_n\xi_n(kr)\pi_n(\theta)\sin\theta\cos\varphi, \label{eq:Esca_r}\\
    \mathcal{E}^{{\rm sca}\,(n)}_{\theta} &=& -i^n\frac{2n+1}{kr(n+1)n}\bigg[b_n\xi_n(kr)\pi_n(\theta)- \label{eq:Esca_theta}\\
    & &-ia_n\xi'_n(kr)\tau_n(\theta)\bigg]\cos\varphi, \nonumber\\
        \mathcal{E}^{{\rm sca}\,(n)}_{\varphi} &=& i^n\frac{2n+1}{kr(n+1)n}\bigg[b_n\xi_n(kr)\tau_n(\theta)-\label{eq:Esca_phi}\\
    & &-ia_n\xi'_n(kr)\pi_n(\theta)\bigg]\sin\varphi, \nonumber\\
& &\mbox{}\nonumber\\
        \mathcal{H}^{{\rm sca}\,(n)}_{r} &=& i^{n+1}\frac{2n+1}{(kr)^2}b_n\xi_n(kr)\pi_n(\theta)\sin\theta\sin\varphi, \label{eq:Hsca_r}\\
          \mathcal{H}^{{\rm sca}\,(n)}_{\theta} &=& -i^n\frac{2n+1}{kr(n+1)n}\bigg[a_n\xi_n(kr)\pi_n(\theta)- \label{eq:Hsca_theta}
\end{eqnarray}
\begin{eqnarray}
    & &-ib_n\xi'_n(kr)\tau_n(\theta)\bigg]\sin\varphi, \nonumber\\
    \mathcal{H}^{{\rm sca}\,(n)}_{\varphi} &=&
    -i^n\frac{2n+1}{kr(n+1)n}\bigg[a_n\xi_n(kr)\tau_n(\theta)-\label{eq:Hsca_phi}\\
    & &-ib_n\xi'_n(kr)\pi_n(\theta)\bigg]\cos\varphi. \nonumber\\
 & &\mbox{}\nonumber\\
     \mathcal{E}^{{\rm in}\,(n)}_{r} &=&\! i^{n-1}\!\frac{2n+1}{(mkr)^2}d_n\psi_n(mkr)\pi_n(\theta)\!\sin\theta\!\cos\varphi,\; \label{eq:Ein_r}\\
      \mathcal{E}^{{\rm in}\,(n)}_{\theta} &=& i^n\frac{2n+1}{mkr(n+1)n}\bigg[c_n\psi_n(mkr)\pi_n(\theta)- \label{eq:Ein_theta}\\
    & &-id_n\psi'_n(mkr)\tau_n(\theta)\bigg]\cos\varphi, \nonumber\\
    \mathcal{E}^{{\rm in}\,(n)}_{\varphi} &=& -i^n\frac{2n+1}{mkr(n+1)n}\bigg[c_n\psi_n(mkr)\tau_n(\theta)-\label{eq:Ein_phi}\\
    & &-id_n\psi'_n(mkr)\pi_n(\theta)\bigg]\sin\varphi, \nonumber
\end{eqnarray}
\begin{eqnarray}
    \mathcal{H}^{{\rm in}\,(n)}_{r} &=&\! i^{n-1}\frac{2n+1}{m(kr)^2}c_n\psi_n(mkr)\pi_n(\theta)\!\sin\theta\!\sin\varphi, \label{eq:Hin_r}\\
    \mathcal{H}^{{\rm in}\,(n)}_{\theta} &=& i^n\frac{2n+1}{kr(n+1)n}\bigg[d_n\psi_n(mkr)\pi_n(\theta)- \label{eq:Hin_theta}\\
    & &-ic_n\psi'_n(mkr)\tau_n(\theta)\bigg]\sin\varphi, \nonumber\\
    \mathcal{H}^{{\rm in}\,(n)}_{\varphi} &=& i^n\frac{2n+1}{kr(n+1)n}\bigg[d_n\psi_n(kr)\tau_n(\theta)-\label{eq:Hin_phi}\\
    & &-ic_n\psi'_n(mkr)\pi_n(\theta)\bigg]\cos\varphi. \nonumber
\end{eqnarray}
Здесь $\xi(z)$ связана с функцией Ганкеля первого рода $H^{(1)}_{\nu}$ соотношением $\xi_n(z) = \sqrt{\frac{\pi z}{2}}H^{(1)}_{n+\frac{1}{2}}(z)$, а коэффициенты рассеяния $a_n,\;b_n,\;c_n$ и $d_n$ определяются из условия непрерывности тангенциальных компонент полей $\mathbf{E}$ и $\mathbf{H}$ на поверхности сферы, что приводит к следующим их значениям:

\begin{eqnarray}
  a_n &=& \frac{m\psi_n(mx)\psi_n'(x) - \psi_n(x)\psi_n'(mx)}{m\psi_n(mx)\xi_n'(x) - \xi_n(x)\psi_n'(mx)}, \label{eq:a_n} \\
  b_n &=& \frac{m\psi_n(x)\psi'_n(mx)-\psi_n(mx)\psi'_n(x)}{m\xi_n(x)\psi'_n(mx)-\psi_n(mx)\xi'_n(x)}, \label{eq:b_n} \\
  c_n &=& -\frac{im}{m\xi_n(x)\psi'_n(mx)-\psi_n(mx)\xi'_n(x)}, \label{eq:c_n}\\
  d_n &=& \frac{im}{m\psi_n(mx)\xi_n'(x) - \xi_n(x)\psi_n'(mx)}\label{eq:d_n}
\end{eqnarray}
(при написании \eqref{eq:c_n}--\eqref{eq:d_n} нами использовалось тождество $ \psi_n(x)\xi'_n(x) - \psi'_n(x)\xi_n(x) \equiv i$~\cite{Abramowitz::1965}).

Отметим, что, как следует из приведенных выражений, радиальные компоненты полей рассеянных парциальных сферических волн не равны нулю. Это означает, что такие волны не вполне поперечны. Они становятся поперечными только асимптотически при $r \rightarrow \infty$, т.к. радиальные компоненты с ростом $r$ убывают быстрее, чем тангенциальные.

Зная координатные зависимости всех компонент рассеянного поля, и, вычисляя векторное произведение $\mathbf{E}$ на $\mathbf{H}$, можно получить профиль вектора Пойнтинга, из которого, в частности, следует выражения~\eqref{eq:Ipar},~\eqref{eq:Iper}. Мы не будем здесь этим заниматься.

Укажем также, что аналогичное точное решение может быть получено для задачи о рассеянии плоской линейно поляризованной волны бесконечным круговым цилиндром. При этом его удается получить при произвольной ориентации оси цилиндра относительно волнового вектора падающей волны и плоскости ее поляризации. Отличие от задачи для сферы состоит в том, что, естественной координатной системой теперь является цилиндрическая, и разложение следует проводить не по сферическим, а по цилиндрическим функциям. Однако явный вид такого решения весьма громоздкий и здесь приводится не будет. Его можно найти, например в~\cite{Kerker2013,Bohren::1998}.

\subsection{Общие свойства коэффициентов рассеяния}

Из выражений \eqref{eq:Esca_r}--\eqref{eq:Hsca_phi}, \eqref{eq:a_n}, \eqref{eq:b_n} следует, что вся зависимость рассеянного излучения от индивидуальных свойств рассеивающей частицы определяется коэффициентами рассеяния $a_n,\;b_n$ --- все остальные, входящие в \eqref{eq:Esca_r}--\eqref{eq:Hsca_phi} выражения универсальны для любых сферических частиц. Через эти же коэффициенты рассеяния выражаются и все парциальные сечения, \mbox{см. \eqref{eq:sigma_sph_n}, \eqref{eq:sigma_sph_abs_n}.} Аналогичные утверждения справедливы и при рассеянии света цилиндром. По этой причине коэффициенты рассеяния играют в рассматриваемой задаче принципиальную роль. Покажем, что эти коэффициенты имеют некоторые важные универсальные свойства, следующие из общих свойств задачи рассеяния, и скрытые в громоздких выражениях \eqref{eq:a_n}, \eqref{eq:b_n}~\cite{Tribelsky:2013ft,PhysRevLett.120.033902_2018-UltimateAbsorption}.

Для выяснения этих свойств воспользуемся тем, что ортогональные друг другу моды, входящих в мультипольные разложения рассеянного поля, являются собственными функциями соответствующей краевой задачи. Это означает, что каждая из них не имеет проекций ни на какую другую и может возбуждаться независимо. То же справедливо и для электрических и магнитных мод с одним и тем же значением $n$, т.к. они не зацеплены друг с другом через граничные условия, а задача рассеяния линейна.

Далее, отметим, что формулы \eqref{eq:sigma_sph_n}, \eqref{eq:sigma_sph_abs_n} для парциальных сечений непосредственно вытекают из ортогональности собственных функций задачи и оптической теоремы. Последняя, в свою очередь, является прямым следствием закона сохранения энергии~\cite{Kerker2013,Bohren::1998,Born_Wolf:Optics}. Поэтому они имеют общность значительно большую, чем обсуждаемое сейчас решение задачи Ми для сферы.

Рассмотрим теперь возбуждение такой единственной парциальной моды. Соответствующий коэффициент рассеяния обозначим $z_n$, где в качестве $z_n$ может фигурировать как $a_n$, так и $b_n$. Обсудим сначала недиссипативный случай (Im$\,\varepsilon = 0$). При отсутствии диссипации все парциальные сечения поглощения тождественно равны нулю. Будучи применено к выражению \eqref{eq:sigma_sph_abs_n}, данное условие приводит к равенству Re$\,z_n = |z_n|^2$. Представляя комплексный коэффициент рассеяния в виде $z_n = z_n^\prime + i z_n^{\prime\prime}$, убеждаемся, что это эквивалентно условию \mbox{$z_n^{\prime} - z_n^{\prime\prime 2} = z_n^{\prime 2} \geq 0$}. Отсюда немедленно следует, что \mbox{$z_n^{\prime} \geq z_n^{\prime\prime 2} \geq 0$} и \mbox{$z_n^{\prime} \geq z_n^{\prime 2}$}, что, в свою очередь, приводит к ограничению \mbox{$0 \leq z_n^{\prime} \leq 1$}. Наконец, любое положительное число, не превосходящее единицы, без нарушения общности можно представить в виде
\begin{equation}\label{eq:an'_FG}
  z_n^\prime = \frac{F_n^2}{F_n^2+G_n^2},
\end{equation}
где, $F_n$ и $G_n$ --- действительные величины. Воспользовавшись представлением \eqref{eq:an'_FG} и равенством \mbox{$z_n^{\prime\prime 2} = z_n^{\prime} - z_n^{\prime 2}$}, находим, что в этом случае
\begin{equation}\label{eq:an"_FG}
  z_n^{\prime\prime} = \pm\frac{F_nG_n}{F_n^2+G_n^2}.
\end{equation}

Поскольку знаки $F$ и $G$ еще не определены, выбор конкретного знака в правой части \eqref{eq:an"_FG} не нарушает общности рассмотрения. Для определенности выберем знак минус. Тогда, собирая все вместе, получаем

\begin{equation}\label{eq:an_full_FG}
  z_n=\frac{F_n^2}{F_n^2+G_n^2}-i \frac{F_nG_n}{F_n^2+G_n^2} \equiv \frac{F_n}{F_n+iG_n},
\end{equation}
где действительные $F$ и $G$, разумеется, являются функциями $\varepsilon$ и $R$ (фактически $m$ и $x$).

Что будет при конечной диссипации, когда \mbox{$\varepsilon = \varepsilon^\prime + i\varepsilon^{\prime\prime}$} и $\varepsilon^{\prime\prime} \neq 0$? Чтобы ответить на этот вопрос заметим, что ни уравнения Максвелла, ни соответствующие граничные условия не содержат членов, в которых $\varepsilon$ входила бы в комбинации, отличной от $\varepsilon^\prime + i\varepsilon^{\prime\prime}$. Это означает, что переход к конечной диссипации сводится к формальной замене действительного $\varepsilon$ на комплексное. Такая замена делает функции $F$ и $G$ комплексными, но не меняет \emph{структуры\/} \mbox{выражения \eqref{eq:an_full_FG}}. Действительно, воспользовавшись известным представлением функции Ганкеля: \mbox{$ H^{(1)}_\nu(z) \equiv J_\nu(z) + iY_\nu(z)$}, где $Y_\nu(z)$ --- функция Неймана~\cite{Abramowitz::1965}, можно убедиться, что выражения \eqref{eq:a_n}, \eqref{eq:b_n} в самом деле соответствуют \eqref{eq:an_full_FG} в том числе и при \mbox{комплексном $\varepsilon$.}
При этом $F_n$ равны числителям правых частей выражений \eqref{eq:a_n}, \eqref{eq:b_n}, а соответствующие $G_n$ определяются выражениями
\begin{eqnarray}
  G_n^{\,{\rm sph}\,(a)} &=& \psi_n^\prime(mx)\chi_n(x)- m\psi_n(mx)\chi^\prime_n(x), \label{eq:Gn_sph_a} \\
  G_n^{\,{\rm sph}\,(b)}  &=& \psi_n(mx)\chi^\prime_n(x)- m\psi^\prime_n(mx)\chi_n(x), \label{eq:Gn_sph_b}
\end{eqnarray}
где $\chi_n(x) = - \sqrt{\frac{\pi x}{2}}Y_{n+\frac{1}{2}}(x)$.

Запись коэффициентов рассеяния в виде \eqref{eq:an_full_FG} позволяет сделать важный вывод о некоторых их общих свойствах. Как отмечалось в основном тексте обзора, резонансам Ми отвечает условие $G_n=0$. В этом случае соответствующий коэффициент рассеяния обращается в единицу, а парциальное сечение рассеяния достигает своего максимума. Как было показано, при отсутствии диссипации все $F_n$ и $G_n$ чисто действительны. Тогда выражение \eqref{eq:an_full_FG} можно переписать в виде \mbox{$z_n(\zeta) =({\rm Re}\,\zeta)/\zeta$,} где \mbox{$\zeta \equiv F_n + iG_n$.}

Существенно, что определенная таким образом функция комплексной переменной $z_n(\zeta)$ \emph{не является аналитической ни в одной точке комплексной плоскости $\zeta$.} В этом легко убедиться, например, применив к ней условия Коши-Римана. Если бы $z_n(\zeta)$ была аналитической функцией, то согласно этим условиям должно было бы выполняться тождество
\begin{equation}\label{eq:Cauchy_Riemann}
  \frac{\partial z_n}{\partial F_n}+i \frac{\partial z_n}{\partial G_n} \equiv 0.
\end{equation}
Легко видеть, что в действительности левая часть выражения \eqref{eq:Cauchy_Riemann} равна $1/(F_n+iG_n) \neq 0$, и условия Коши-Римана не выполняются.

Точка $\zeta=0$ является особой точкой этой функции, \emph{в которой она не имеет никакого определенного значения\/} --- значение $z_n(\zeta)$ в этой точке зависит от траекторий на плоскости $\zeta$ по которой мы к ней приближаемся. Действительно, рассмотрим семейство траекторий вида $|G_n|=A|F_n|^\alpha$ где $A$ и $\alpha$ положительные числа. Видно, что в этом случае при $\zeta \rightarrow 0$ функция \mbox{$z_n \rightarrow 0$} при \mbox{$0<\alpha <1$;} \mbox{$z_n \rightarrow 1/(1\pm iA)$} при  \mbox{$0<\alpha <1$,} где знак плюс берется если $F_n$ и $G_n$ имеют одинаковые знаки и минус, если противоположные. И наконец, \mbox{$z_n \rightarrow 1$} при  \mbox{$\alpha >1$}.

Чтобы понять, почему такая простая функция обладает столь, на первый взгляд, странными свойствами, рассмотрим график $|z_n(\zeta)|^2$, который представляет собой определенную поверхность в трехмерном пространстве, см. Рис.~\ref{fig:zn2}. Видно, что два крыла этой поверхности пересекаются по линии $\zeta=0$, что и объясняет отмеченные выше свойства. Из сказанного ясно, что особенность такого типа является общим свойством коэффициентов рассеяния и, как будет видно из дальнейшего, играет важную роль в понимании различных резонансных эффектов при рассеянии света субволновыми частицами с малой диссипацией.

В частности, зависимость значения коэффициента рассеяния в точке $\zeta = 0$ от траектории, по которой к этой точке приближаются, делает принципиально важным правильный выбор такой траектории. Если в математическом смысле все возможные траектории эквивалентны, то в соответствующей физической задаче естественным является набор переменных $R,\;\omega$. При этом физически осмысленной постановкой задачи является изучение рассеяния частицей с данным значением $R$ в зависимости от частоты падающего излучения. Как отмечалось в основном тексте обзора,
ширина резонансных линий резко уменьшается с уменьшением размера частицы. Поэтому при таком подходе и достаточно малом $R$ становится необходимым выйти за рамки приближения монохроматической падающей волны и рассмотреть конечную ширину линии источника. Последовательное применение этого подхода снимает математическую неопределенность в значениях коэффициентов рассеяния в точке $\zeta = 0$. Более подробно этот вопрос обсуждается в работе~\cite{Brynkin2019}.

Учет сколь угодно малой, но конечной диссипации смещает значения резонансных частот (решения уравнения $G_n=0$) с действительной оси в комплексную плоскость. В результате $G_n$ не обращается в ноль ни при каких действительных значениях $\omega$, и особенность при $\zeta = 0$ исчезает. {В этом случае при чисто действительных $\omega$ функция $G_n$ становится комплексной: \mbox{$G_n=G_n'+iG_n''$}. При малой диссипации (\mbox{$|G_n'| \gg G_n''$}) профиль $|z_n(\zeta)|^2$ после соответствующего перемасштабирования переменных приобретает универсальный вид, аналогичный выражению \eqref{eq:kappa(xi,zeta)}:
\begin{equation}\label{eq:zn2_dissip}
  |z_n|^2 \approx \frac{\zeta'^2}{(\zeta'+1)^2+\zeta''^2},\; \zeta = -\frac{F_n'}{G_n''}+i\frac{G_n'}{G_n''},
\end{equation}
см. Рис.~\ref{fig:zn2_diss}.   }
Однако, несмотря на устранение особенности, все сказанное выше относительно резонансного рассеяния в окрестности точки $R=0$ остается справедливым и при конечной диссипации до тех пор пока радиационное затухание преобладает над диссипативным~\cite{Brynkin2019}.

\begin{figure}
  \centering
  \includegraphics[width=\columnwidth]{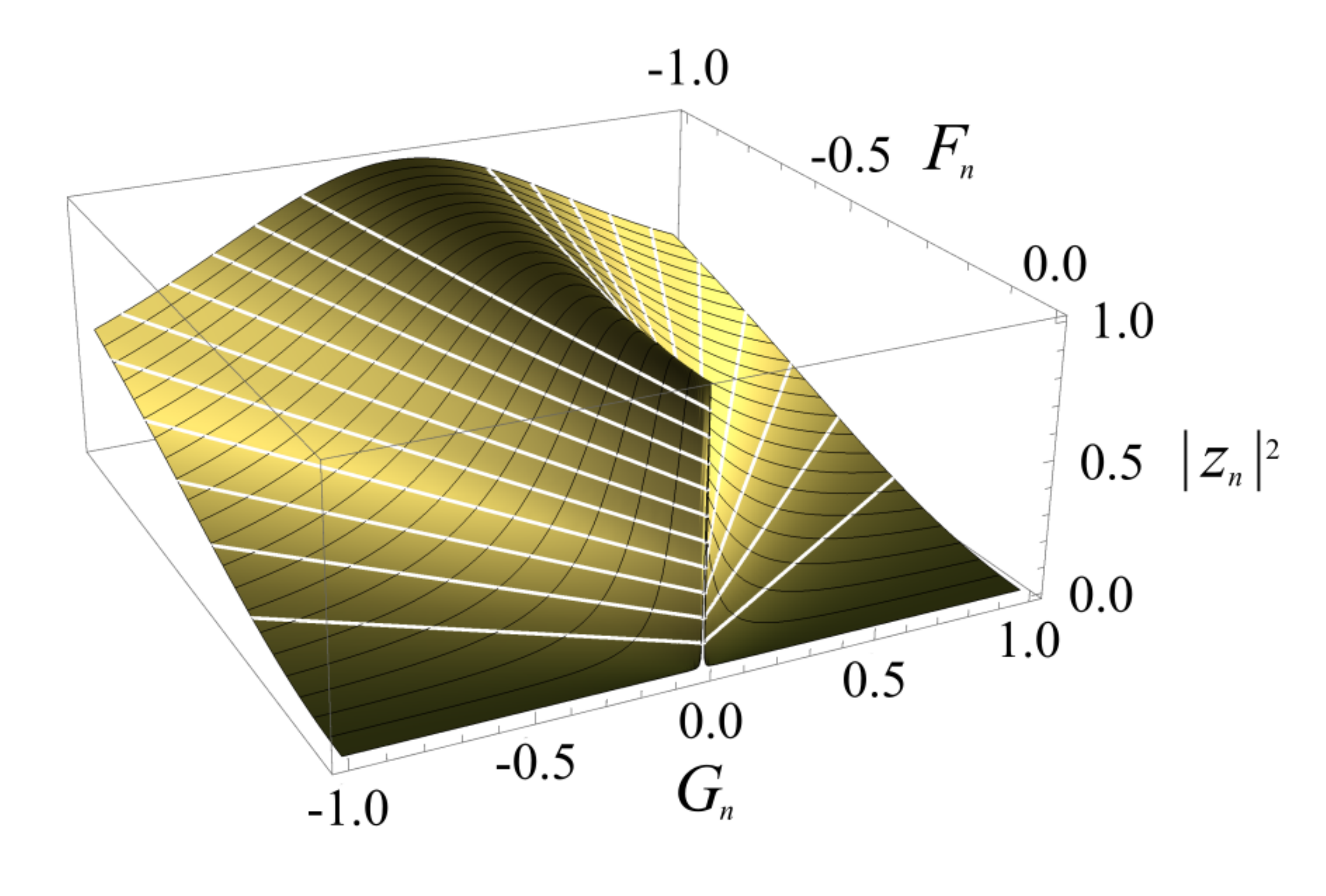}
  \caption{Поверхность $|z_n(\zeta)|^2$ {в соответствии с выражением \eqref{eq:an_full_FG} при чисто действительных $F_n$ и $G_n$ (недиссипативный предел)} и семейство линий $F_n = const$ на ней (тонкие черные линии). Чем ближе $F_n$ к нулю, тем круче максимум линий при $G_n=0$. В конце концов, два крыла поверхность пересекаются вдоль линии $F_n=G_n=0$. Хотя различные траектории, лежащие на поверхности, могут пересекать эту линию при различные значениях $|z_n|^2$, все точки пересечения проецируются в одну и ту же точку $F_n=G_n=0$ на плоскости $(F_n,G_n)$. Как пример этих траекторий, набор линий уровня $G_n = const$ для разных значений константы показаны толстыми белыми линиями~\cite{Brynkin2019}.}\label{fig:zn2}
\end{figure}
\begin{figure}
  \centering
  \includegraphics[width=\columnwidth]{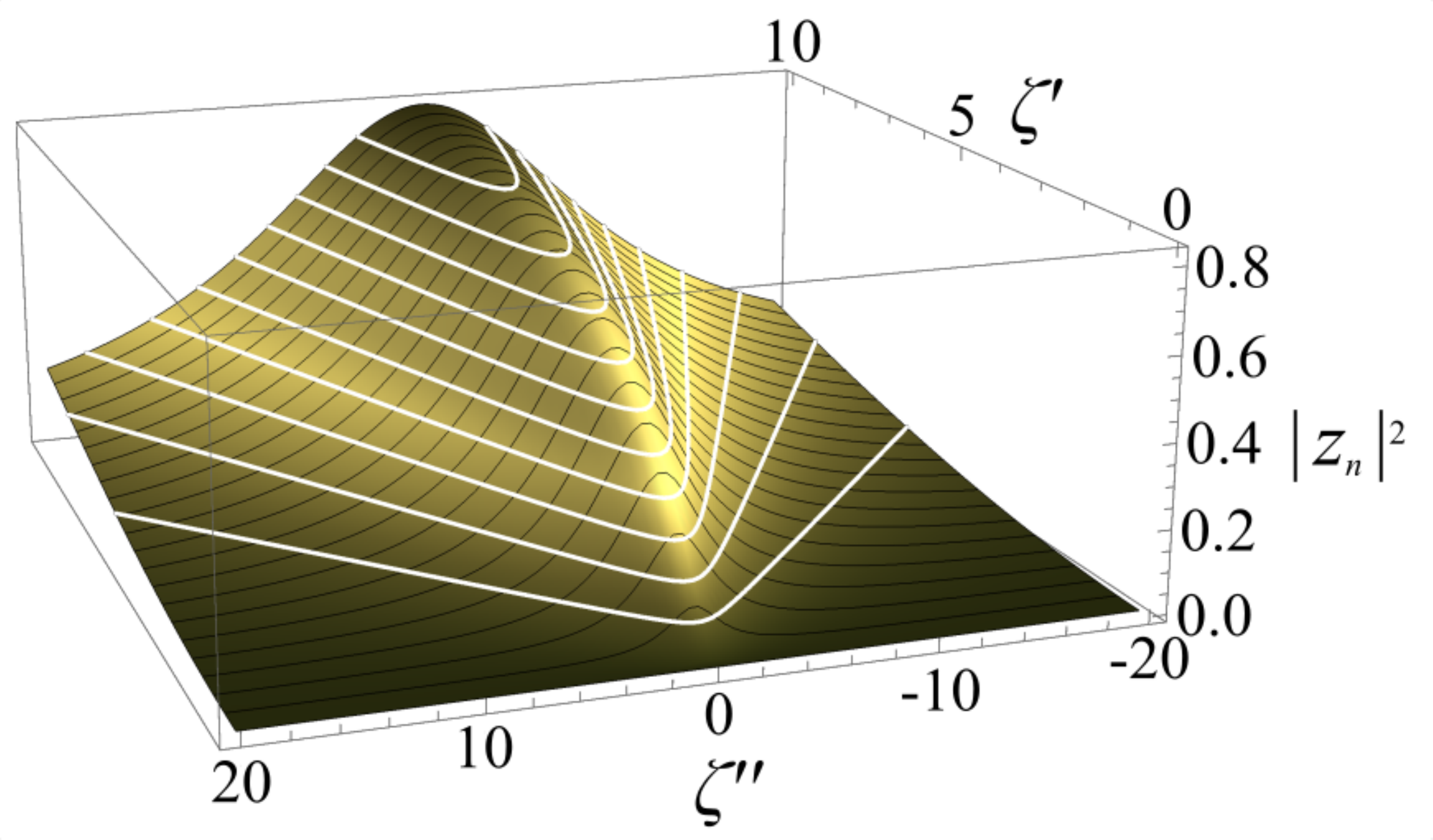}
  \caption{То же, что на Рис.~\ref{fig:zn2}, но с учетом малой диссипации, см. \eqref{eq:zn2_dissip}.}\label{fig:zn2_diss}
\end{figure}

В заключение этого обсуждения отметим, что неаналитичность коэффициентов рассеяния в недиссипативном пределе проявляется в представлении их как функции $\zeta$. Если вместо этого рассматривать их как функции комплексного параметра размера $x$ (или комплексного коэффициента преломления $m$), то в силу представления $F_n$ и $G_n$ через функции Бесселя, см. \mbox{\eqref{eq:a_n}--\eqref{eq:c_n},} \mbox{\eqref{eq:Gn_sph_a}--\eqref{eq:Gn_sph_b}} и аналитических свойств последних, все они оказываются аналитическими функциями $x$ (или $m$) на всей комплексной плоскости за исключением счетного множества особых точек.

\bibliography{UFN} 
\end{document}